\documentclass[a4paper,12pt]{iopart}
\usepackage[utf8]{inputenc}
\usepackage[hidelinks]{hyperref}
\usepackage{color}
\usepackage{tensor}
\usepackage{ulem}
\usepackage{iopams}
\eqnobysec 
\newcommand{\eqend}[1]{\,\mathrm{#1}}

\newcommand{\fpartial}[2]{\frac{ \partial #1}{\partial #2}}

\newcommand{\bra}[1]{{\left\langle{#1}\right\vert}}
\newcommand{\ket}[1]{{\left\vert{#1}\right\rangle}}

\newcommand{\D}{\mathrm{d}}
\newcommand{\Dt}{\mathrm{d}t}
\newcommand{\Ds}{\mathrm{d}s}
\newcommand{\ee}{\mathrm{e}}
\newcommand{\ii}{\mathrm{i}}
\newcommand{\beq}{\begin{equation}}
\newcommand{\eeq}{\end{equation}}
\newcommand{\beqa}{\begin{eqnarray}}
\newcommand{\eeqa}{\end{eqnarray}}

\usepackage{tikz}
\usetikzlibrary{decorations.markings}

\usetikzlibrary{arrows.meta}
\tikzset{>=latex}
\tikzset{->-/.style={decoration={
			markings,
			mark=at position #1 with {\arrow{>}}},postaction={decorate}}}

\begin{document}

\title[Automorphic scalar fields in $\mathrm{dS}_2$]{Automorphic scalar fields in two-dimensional de~Sitter space}

\author{Atsushi~Higuchi\(^1\), Lasse~Schmieding\(^1\) and David~Serrano~Blanco\(^1\)}
\address{\(^1\) Department of Mathematics, University of York, Heslington, York, YO10 5DD, United Kingdom}

\eads{\mailto{atsushi.higuchi@york.ac.uk}, \mailto{lasse.schmieding@gmail.com}, \mailto{david\_serranob@outlook.com} }

\begin{abstract}
We study non-interacting automorphic quantum scalar fields with positive mass in two-dimensional de~Sitter space.  We find that there are no Hadamard states which are de~Sitter invariant except in the periodic case, extending the result of Epstein and Moschella for the anti-periodic case.  We construct the two-point Wightman functions for the non-Hadamard de~Sitter-invariant states by exploiting the fact that they are functions of the geodesic distance between the two points satisfying an ordinary differential equation.  We then examine a certain Hadamard state, which is not de~Sitter invariant, and show that it is approximately a thermal state with the Gibbons-Hawking temperature when restricted to a static region of the spacetime.
\end{abstract}

\maketitle

\section{Introduction}\label{sec:introduction}

Unlike higher-dimensional de~Sitter spaces, the two-dimensional de~Sitter space, $\mathrm{dS}_2$, is not simply connected. Consequently, the behaviour of the fields under complete traversals of non-contractible loops can be made non-trivial. This topological non-triviality can have interesting consequences. For Dirac spinor fields in $\mathrm{dS}_2$, Epstein and Moschella found that an anti-periodic Neveu-Schwarz boundary condition is more natural than a periodic Ramond boundary condition~\cite{epstein2016sitter}. On conformally mapping the spinors from a flat timelike cylinder to $\mathrm{dS}_2$, they found that only the anti-periodic spinor fields possess a form of invariance under all de~Sitter transformations. Furthermore, for free scalar fields on $\mathrm{dS}_2$, they showed that the properties of the anti-periodic ones is quite different from the periodic ones~\cite{epstein2018topological,epstein2020annales}. For masses that would correspond to unitary representations of the symmetry group $\mathrm{SL}(2,\mathbb{R})$ of the spacetime in the complementary series for the periodic case, the anti-periodic fields do not admit de~Sitter-invariant two-point functions. This can be understood from the representation theory of $\mathrm{SL}(2, \mathbb{R})$: there are no unitary irreducible representations corresponding to this mass range for the anti-periodic scalar fields. They also showed that for the anti-periodic case there is no natural analogue of the Bunch-Davies vacuum state~\cite{schomblond1976dS, bunch1978quantum, chernikov1968quantum, allen1985vacuum} for any value of the mass.

In fact the non-trivial fundamental group $\pi_1(\mathrm{dS}_2) = \mathbb{Z}$ of $\mathrm{dS}_2$ allows for automorphic scalar fields~\cite{avis1978vacuum, isham1978twisted, banach1979vacuum, Banach1979automorphic, banach1980quantum}, which include the anti-periodic ones. The automorphic scalar fields are generically complex scalar fields, which transform under a unitary representation of $\pi_1(\mathrm{dS}_2)$ on traversal of the non-contractible loop. 

A coordinate system of $\mathrm{dS}_2$ can be chosen so that the time variable $t$ runs from $-\infty$ to $\infty$ whereas the space coordinate $\phi$ is an angular variable with the identification $\phi \sim \phi+2\pi$.  An automorphic scalar field $\Phi(t,\phi)$ satisfies the following periodicity condition:
	\begin{equation}
		\Phi(t, \phi + 2 \pi) = \ee^{2 \pi \ii \beta} \Phi(t, \phi),
	\end{equation}
where $- 1/2 < \beta \leq 1/2$. This field can naturally be viewed as a single-valued field on the universal covering space, $\widetilde{\mathrm{dS}}_2$, of $\mathrm{dS}_2$, and transforms under representations of $\widetilde{\mathrm{SL}}(2, \mathbb{R})$, the universal covering group of (the component of the identity of) the de~Sitter symmetry group $\mathrm{SO}_0(2,1)$. 

In this paper we investigate the non-interacting quantum automorphic scalar field in $\mathrm{dS}_2$  with an arbitrary value of $\beta$, extending the work of Epstein and Moschella.
We show that de~Sitter-invariant states are inevitably non-Hadamard for any nonzero value of $\beta$.  We then construct the two-point Wightman functions for these de~Sitter-invariant states by exploiting the fact that they are functions of the geodesic distance between the two points which satisfy the Legendre equation.  We also study a de~Sitter non-invariant Hadamard state in detail and show that it approximately exhibits the Gibbons-Hawking effect~\cite{gibbons1977cosmological}, i.e.\ that the reduced state obtained by restricting it to a static patch of this spacetime shows features of an approximate thermal state with the temperature $H/(2\pi)$, where $H$ is the Hubble constant of the spacetime.  This clarifies the extent of the ``disappearance of the Gibbons-Hawking effect'', 
which was observed by Epstein and Moschella for anti-periodic scalar fields~\cite{epstein2020annales}.

The rest of the paper is organised as follows. In section~\ref{sec:geometry} we review the geometry of $\mathrm{dS}_2$ and its covering space, $\widetilde{\mathrm{dS}}_2$. In section~\ref{sec:dS2scalars} we introduce the automorphic scalar fields and provide mode expansions for the corresponding quantum fields in global coordinates. In section~\ref{sec:dSinvariance} we study the de~Sitter-invariance properties of the Fock vacuum states for the automorphic scalar fields
and find that if there is a unitary representation of $\widetilde{\mathrm{SL}}(2, \mathbb{R})$ with the periodicity parameter $\beta$ and the mass $M$ in the principal or complementary series, then there are de~Sitter-invariant vacuum states with two parameters, which correspond to the $\alpha$-vacua in the periodic case~\cite{mottola1985particle,allen1985vacuum}.
In section~\ref{sec:hadamard} we show that there is a Hadamard state among these de~Sitter-invariant states if and only if $\beta=0$ (the periodic case). In section~\ref{sec:explicit-2p-functions} we construct the Wightman two-point functions for the de~Sitter-invariant states found in section~\ref{sec:dSinvariance}.
In section~\ref{sec:noninvhad} we define a natural de~Sitter non-invariant state for each periodicity parameter $\beta$ and mass $M$ and prove that it is Hadamard.
Then, in section~\ref{sec:gibbons-hawking-non-inv} we show that this state approximately exhibits the Gibbons-Hawking effect.
In section~\ref{sec:discussion} we summarise and discuss our results. 
In \ref{Appendix:S-beta} we find conditions on the rotationally-invariant vacuum states.
In \ref{Appendix:formula-for-Pl} we prove a transformation formula for the Legendre functions of the first kind.
In \ref{appendix-integral} we present the proof of an integral representation of a series used in this paper. In \ref{appendix-GH-effect} we review the Gibbons-Hawking effect for the periodic case. In \ref{appendix:bound-Ferrers}  we establish a bound on an integral of a Ferrers function, which is used in section~\ref{sec:gibbons-hawking-non-inv}. 

\section{Geometry of two-dimensional de~Sitter space}\label{sec:geometry}
	
Two-dimensional de~Sitter space can be realised as a hyperboloid embedded within three-dimensional Minkowski space. Let $X^0$, $X^1$ and $X^2$ be coordinates in the three-dimensional embedding space with Minkowski metric,
	\begin{equation}\label{eq:flat-metric}
		\mathrm{d} s^2 = - (\mathrm{d} X^0)^2 + (\mathrm{d} X^1)^2 + (\mathrm{d} X^2)^2 \eqend{.}
	\end{equation}
Then the de~Sitter hyperboloid is determined by the equation
	\begin{equation}\label{eq:hyperboloid}
             -(X^0)^2 + (X^1)^2 + (X^2)^2 = H^{-2}\,,
	\end{equation}
where $H$ is a positive constant.  We let $H=1$ from now on.
The metric of $\mathrm{dS}_2$ is obtained by restricting the flat metric~(\ref{eq:flat-metric}) to the hyperboloid. The symmetries of the hyperboloid are Lorentz transformations of the embedding space, and the de~Sitter metric therefore inherits an $\mathrm{SO}(2,1)$ symmetry group.

The hyperboloid given by (\ref{eq:hyperboloid}) (with $H=1$) can be parametrised as
	\begin{equation}\label{eq:dS2global}
		\eqalign{X^0 &= \sinh t \eqend{,} \\
		X^1 &= \cosh t \cos \phi \eqend{,} \\
		X^2 &= \cosh t \sin \phi \eqend{,} }
	\end{equation}
where $- \infty < t < \infty$ and $\phi$ is a periodic variable identified as $\phi\sim \phi + 2\pi$.  These coordinates cover the whole de~Sitter space, and
the metric in these coordinates takes the following form:
	\begin{equation}\label{eq:dS2globalmetric}
		\mathrm{d} s^2 = - \mathrm{d} t^2 + \cosh^2 t \ \mathrm{d} \phi^2 \eqend{.}
\end{equation}
By introducing conformal coordinates $(\tau, \phi)$, related to the global coordinates by
\begin{equation}\label{eq:conformal-coordinates}
    \sinh t = \tan \tau\eqend{,}
\end{equation} 
with $- \pi/2 < \tau < \pi/2$, the conformal flatness of de~Sitter space is made manifest. The metric in conformal coordinates takes the form
	\begin{equation}\label{eq:dS2conformalmetric}
		\mathrm{d} s^2 = \sec^2 \tau \left( - \mathrm{d} \tau^2 + \mathrm{d} \phi^2 \right) \eqend{.}
	\end{equation}

Given two points $x$ and $x'$ in $\mathrm{dS}_2$, we can use the embedding space to define a de~Sitter-invariant product $Z(x,x')$ between them. This is given by
	\begin{equation}
		Z(x,x') = - X^0(x) X^0(x') + X^1(x) X^1(x') + X^2(x) X^2(x') \eqend{,}
	\end{equation}
where $X^A(x)$ are the embedding-space coordinates of the point $x$. In terms of global and conformal coordinates $x = (t,\phi) = (\tau, \phi)$ and $x' = (t', \phi') = (\tau', \phi')$ we have
	\begin{equation}\label{eq:defZ}
		\eqalign{ Z(x,x') &= - \sinh t \sinh t' + \cosh t \cosh t' \cos (\phi - \phi') \\ 
		&= \sec \tau \sec \tau' \left[ - \sin \tau \sin \tau' + \cos (\phi - \phi') \right] \eqend{.} }
	\end{equation}
For points connected by a geodesic, it is possible to relate $Z(x,x')$ to the geodesic distance $\mu(x,x')$ between the points as $Z(x,x') = \cos \mu(x,x')$. [We let $\mu(x,x')$ be purely imaginary if $x$ and $x'$ are timelike separated.] We use this relation to continue $\cos \mu(x,x')$ to points which are not connected by a geodesic.
	\begin{figure}[t]\label{fig:CarterPenrose}
\begin{center}
	\begin{tikzpicture}

		\draw (-4,2) -- (4,2) (4,-2) -- (-4,-2);
		\draw[dashed] (-4,-2) -- (-4,2) (4,2) -- (4,-2);

		\draw[dashed, opacity = 0.5] (-2,-2) -- (2,2) (-2,2) -- (2,-2);
		\draw[dashed, opacity = 0.5] (-4,0) -- (-2,2) (-4,0) -- (-2,-2) (4,0) -- (2,2) (4,0) -- (2,-2);
		
		\filldraw (0,0) circle (0.05) node[above] {$O$} (-4,0) circle (0.05) (4,0) circle (0.05);		
		
		\filldraw[blue, opacity = 0.1] (-2,2) -- (2,2) -- (0,0) -- (-2,2) (0,0) -- (2,-2) -- (-2,-2) -- (0,0);
		
		\filldraw[red, opacity = 0.1] (-4,0) -- (-4,2) -- (-2,2) -- (-4,0) (-4,0) -- (-2,-2) -- (-4,-2) -- (-4,0) (4,0) -- (4,2) -- (2,2) -- (4,0) (4,0) -- (4,-2) -- (2,-2) -- (4,0);
		
		\filldraw[green, opacity = 0.1] (0,0) -- (2,2) -- (4,0) -- (2,-2) -- (0,0) (0,0) -- (-2,2) -- (-4,0) -- (-2,-2) -- (0,0);
		
		\draw (-4,-2) node[below] {$\phi = - \pi$} (4,-2) node[below] {$\phi = +\pi$};
		
		\draw (4,-2) node[right] {$\tau = - \frac{\pi}{2}$} (4,2) node[right] {$\tau = + \frac{\pi}{2}$};
	\end{tikzpicture}
	
\end{center}
	\caption{Carter-Penrose diagram for two-dimensional de~Sitter space. The green area can be connected to the origin $O$ by spacelike geodesics. The blue area can be connected to the origin by timelike geodesics. The area shaded in red cannot be connected to the origin by a geodesic.}
	\end{figure}
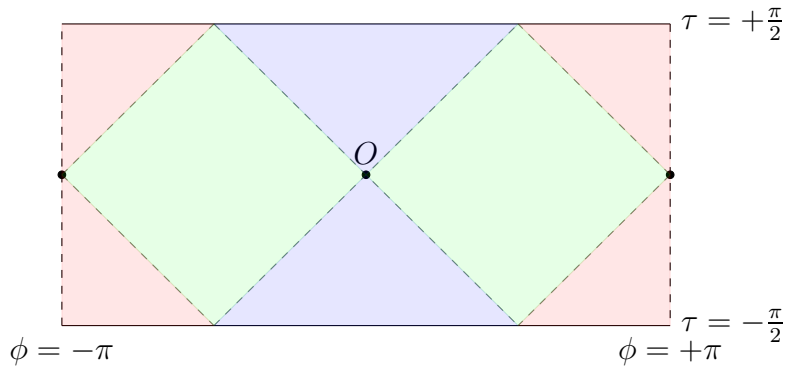	
Another useful coordinate system is the static coordinate system, $(T,R)$, given by
	\begin{equation}\label{eq:dS2static}
		\eqalign{ X^0 &= \sqrt{1 - R^2} \sinh T \eqend{,} \\
		X^1 &= R \eqend{,} \\
		X^2 &= \sqrt{1 - R^2} \cosh T \eqend{,} }
	\end{equation}
where $- \infty < T < \infty$ and $-1 < R < 1$.	Static coordinates only cover the static patch of de~Sitter space defined by $|X^0|< X^2$.
In these coordinates the metric is
	\begin{equation}\label{eq:dS2staticmetric}
		\mathrm{d} s^2 = - (1-R^2) \mathrm{d} T^2 + \frac{\mathrm{d} R^2}{1 - R^2}\eqend{.}
	\end{equation}
	
Three independent Killing vector fields of $\mathrm{dS}_2$ can be written down in global coordinates as
	\begin{equation}\label{eq:dS2-KVF}
	\eqalign{\xi_0 = X^1 \fpartial{}{X^2} - X^2 \fpartial{}{X^1} = \fpartial{}{\phi} \eqend{,} \\
	\xi_1 = X^1 \fpartial{}{X^0} + X^0 \fpartial{}{X^1} = \cos \phi \fpartial{}{t} - \tanh t \sin \phi \fpartial{}{\phi} \eqend{,} \\
	\xi_2 = X^2 \fpartial{}{X^0} + X^0 \fpartial{}{X^2} = \sin \phi \fpartial{}{t} + \tanh t \cos \phi \fpartial{}{\phi} \eqend{.} }
	\end{equation}
In terms of the embedding space the Killing vector $\xi_0$ generates rotations around the $X^0$ axis and $\xi_1$ and $\xi_2$ generate boosts in the $X^1$ and $X^2$ directions, respectively. 
	
\section{Automorphic scalar fields in two-dimensional de~Sitter space with rotationally invariant vacuum state}\label{sec:dS2scalars}

We let $\Phi(x)$ be a massive automorphic scalar field in $\mathrm{dS}_2$. That is, we let $\Phi(x)$ be a complex scalar field which picks up a phase on making a full rotation of the spatial circle. In global coordinates this means
	\begin{equation}\label{eq:automorphy}
		\Phi(t, \phi + 2 \pi) = \ee^{2 \pi \ii \beta} \Phi(t, \phi) \eqend{,}
	\end{equation}
where $-1/2 < \beta \leq 1/2$. This field obeys a Klein-Gordon equation 
	\begin{equation}
		(\Box - M^2) \Phi(t, \phi) = 0 \eqend{,}
	\end{equation}
which in the global coordinates~(\ref{eq:dS2global}) takes the form
	\begin{equation}\label{eq:KG-eq-global}
		- \frac{1}{\cosh t} \fpartial{}{t} \left( \cosh t \fpartial{\Phi}{t} \right) + \frac{1}{\cosh^2 t} \frac{\partial^2 \Phi}{\partial \phi^2} - M^2 \Phi = 0 \eqend{.}  
	\end{equation}
This is a real differential equation and, hence, the complex conjugate of a solution obeys the same equation.  However, it is important to note that, since the boundary condition~(\ref{eq:automorphy}) is complex in general, the complex conjugate solution satisfies a different boundary condition unless $\beta$ is $1$ or $1/2$. Only when $\beta = 0$ (periodic) or $\beta =  1/2$ (anti-periodic) are the boundary conditions real and, hence, the complex conjugate solution satisfies the same boundary condition. Thus, it is possible to consider real scalar fields only in the periodic and anti-periodic cases. Epstein and Moschella~\cite{epstein2018topological,epstein2020annales} considered real scalar fields with $\beta=0, 1/2$. In this paper the field $\Phi(x)$ is a complex field unless otherwise stated.

Quantisation of this field can be achieved through the canonical method described, e.g., in~\cite{birrell1984quantum, wald1994quantum, fulling1989aspects, parker2009quantum}.  On the space of classical solutions to the Klein-Gordon equation~(\ref{eq:KG-eq-global}), we introduce a Klein-Gordon inner product
	\begin{equation}\label{eq:dS2-KG-Product}
		(\Psi_1, \Psi_2)(t) = \ii \int_{0}^{2 \pi} \mathrm{d} \phi \ \cosh t \left[ \Psi_1^* \fpartial{\Psi_2}{t} - \left( \fpartial{\Psi_1^*}{t} \right) \Psi_2 \right] \eqend{,}
	\end{equation}
which is conserved in the sense that $(\Psi_1, \Psi_2)(t_1) = (\Psi_1, \Psi_2)(t_2)$, provided that both $\Psi_1$ and $\Psi_2$ satisfy  the boundary condition~(\ref{eq:automorphy}) and solve the Klein-Gordon equation~(\ref{eq:KG-eq-global}). 
Let $\mathcal{S}_\beta$ be the space of these solutions.
To quantise this scalar field and define the vacuum state, the first step is to find the orthogonal decomposition $\mathcal{S}_\beta = \mathcal{S}^+_\beta\oplus \mathcal{S}^-_\beta$ such that if $\Psi_1 \in \mathcal{S}^+_\beta$ and $\Psi_2\in \mathcal{S}^-_\beta$ are nonzero solutions, then $(\Psi_1,\Psi_1) >0$, $(\Psi_2,\Psi_2) < 0$ and $(\Psi_1,\Psi_2) = 0$.  The space $\mathcal{S}^+_\beta$ serves as the one-particle Hilbert space with which the Fock space is built. 
Since the choice of $\mathcal{S}^+_\beta$ uniquely determines the vacuum state~\cite{birrell1984quantum, wald1994quantum, fulling1989aspects, parker2009quantum}, it is invariant under a symmetry transformation if and only if the subspace $\mathcal{S}^+_\beta$ of solutions is invariant under this transformation.  In the next section we give all de~Sitter-invariant choices of $\mathcal{S}^+_\beta$.

The group of de~Sitter transformations is the universal covering group $\widetilde{\mathrm{SL}}(2,\mathbb{R})$ of $\mathrm{SO}_0(2,1)$, the component of the identity of $\mathrm{SO}(2,1)$.  The corresponding Lie algebra is generated by the Killing vectors $\xi_0$, $\xi_1$ and $\xi_2$ defined by~(\ref{eq:dS2-KVF}). The Lie bracket relations obeyed by these Killing vectors are
	\begin{equation}\label{eq:Killing-commutator}
		[ \xi_0, \xi_1 ] = - \xi_2, \quad [ \xi_0, \xi_2 ] = \xi_1, \quad [ \xi_1, \xi_2 ] = \xi_0 \eqend{,}
	\end{equation}
which define the Lie algebra  $\mathrm{sl}(2,\mathbb{R})$.
The eigenvalue of the quadratic Casimir operator $Q$ defined by
\begin{equation}\label{eq:Q-definition}
		Q = - \xi_0^2 + \xi_1^2 + \xi_2^2 = \frac{\partial^2}{\partial t^2} + \tanh t \fpartial{}{t} - \frac{1}{\cosh^2 t} \frac{\partial^2}{\partial \phi^2} = - \Box \eqend{.}
	\end{equation}
is $-M^2$. It is convenient to let
\begin{equation}\label{eq:msquared-and-l}
M^2 = -l(l+1)\eqend{.}
\end{equation}
Thus, the eigenvalue of the Casimir operator is $l(l+1)$.
Note also that equation~(\ref{eq:automorphy}) implies that the $2\pi$-rotation $R_{2\pi}= \ee^{2\pi \xi_0}$, which is a central element of $\widetilde{SL}(2,\mathbb{R})$, has the value $\ee^{2\pi\ii \beta}$.  

For the subspace $\mathcal{S}_\beta^+$ to be invariant under $\widetilde{SL}(2,\mathbb{R})$ the solutions in $\mathcal{S}_\beta^+$ must form a unitary representation. The condition for the existence of a unitary representation with the eigenvalues $l(l+1)>0$ and $\ee^{2\pi \ii\beta}$ of the Casimir operator and the $2\pi$-rotation, respectively, is given 
by~\cite{pukanszky1964plancherel,kitaev2017notes}
\begin{equation}\label{eq:condition-on-l}
l=-\frac{1}{2}+\ii\lambda\ \textrm{with}\ \lambda > 0\eqend{,}\ \textrm{or}\ -\frac{1}{2} \leq l < -|\beta|\eqend{.}
\end{equation}
(We can also have $l=-1/2-\ii\lambda$ with $\lambda >0$ or $|\beta|-1<l\leq -1/2$, but these are equivalent to the above values because the eigenvalues of the Casimir operator, $l(l+1)$, is invariant under $l \mapsto -l-1$.) 
Hence, the subspace 
$\mathcal{S}_\beta^+$ can be invariant only for these values of $l$.\footnote{For $M^2=0$ ($l=0$) there is no unitary representation and hence no de~Sitter invariant vacuum state unless $\beta=0$. Although there are no invariant vacuum states~\cite{allen1985vacuum,Ford:1977in} for $\beta=0$ either, the existence of a unitary representation in this case allows one to define a de~Sitter-invariant theory~\cite{Kirsten:1993ug,higuchi1991quantum2}.  
See \cite{DeBievre:1998yx} for a somewhat different approach.} We impose the condition (\ref{eq:condition-on-l}) from now on though in this section we only require the vacuum state to be rotationally invariant.  

Next, we discuss the solutions to the field equation~(\ref{eq:KG-eq-global}) of the form
\begin{equation}\label{eq:def-of-calF_m}
f_m(t,\phi)= \mathcal{F}_m(t)\ee^{\ii m\phi}\eqend{,}
\end{equation}
 for each $m\in \beta + \mathbb{Z}$.
 We find the differential equation satisfied by the functions $\mathcal{F}_m$ from the Klein-Gordon equation~(\ref{eq:KG-eq-global}), with $M^2=-l(l+1)$, as
	\begin{equation}\label{eq:assoc-legendre}
		\frac{\mathrm{d}\ }{\mathrm{d} u} \left( (1-u^2) \frac{\mathrm{d} \mathcal{F}_m}{\mathrm{d} u} \right) + \left( l(l+1) - \frac{m^2}{1-u^2} \right) \mathcal{F}_m = 0 \eqend{,}
	\end{equation}
where $u= \ii\sinh t$. This is the associated Legendre equation~\cite[Eq.~8.700]{gradshteyn2014table}. 
Two linearly independent solutions to equation~(\ref{eq:assoc-legendre}) are $\mathsf{P}_l^{-m}(\ii\sinh t)$ and $\mathsf{P}_l^{-m}(-\ii\sinh t)$, where $\mathsf{P}^{-m}_l(z)$ is the Ferrers (or associated Legendre) function of the first kind, if $l\pm m \notin \mathbb{Z}$~\cite[Eq.~8.707]{gradshteyn2014table}, which is the case for the values of $l$ given by~(\ref{eq:condition-on-l}). 
 The Ferrers functions of the first kind are
 expressed in terms of Gauss's hypergeometric function as
	\begin{equation}\label{eq:Pdefinition}
		\mathsf{P}^{-m}_l(z) = \frac{1}{\Gamma(1 + m)} \left( \frac{1 - z}{1+z} \right)^{m/2} F\left( 1+l,-l;1 + m;\frac{1-z}{2} \right) \eqend{.}
	\end{equation}
Then, the functions $\mathsf{P}_l^{-m}(\ii \sinh t)\ee^{\ii m\phi}$ and
$\mathsf{P}_l^{-m}(-\ii\sinh t)\ee^{\ii m\phi}$, $m\in \beta+\mathbb{Z}$,
form a linearly-independent complete set\footnote{The completeness of these solutions follows from the fact that the functions $\ee^{\ii m\phi}$ form a basis for the space of square-integrable functions on $[0,2\pi]$ and that $\mathsf{P}_l^{-m}(\ii\sinh t)$ and
$\mathsf{P}_l^{-m}(-\ii\sinh t)$ are linearly independent solutions to (\ref{eq:assoc-legendre}), which determines the time-dependence of the solutions proportional to $\ee^{\ii m\phi}$. } of solutions to the Klein-Gordon equation (\ref{eq:KG-eq-global}).

Using the Wronskian identity~\cite[Eq.~14.2.3]{NIST:DLMF}
	\begin{equation}\label{eq:Wronskian1}
\fl \qquad	 \mathsf{P}^{-m}_l(u) \frac{d \mathsf{P}^{-m}_l(-u)}{du} - \frac{d \mathsf P^{-m}_l(u)}{d u} \mathsf{P}^{-m}_l(-u) = \frac{2}{\Gamma(m-l) \Gamma(m+l+1) (1 - u^2)} \eqend{,}
\end{equation}
and the identity $\mathsf{P}^{-m}_l(u)^* = \mathsf{P}^{-m}_{l}(u^*)$ for $l \in \mathbb{R}$ and $l \in -1/2 + i \mathbb{R}$, one can evaluate the Klein-Gordon inner product~(\ref{eq:dS2-KG-Product})
for the solutions $\mathsf{P}_l^{-m}(\pm\ii\sinh t)\ee^{\ii m\phi}$. Then, by defining 
	\begin{equation}\label{eq:def-fm}
 \eqalign{		f_m(t, \phi) & = 
\mathcal{F}_m(t) \ee^{\ii  m \phi} \eqend{,} \\
		g_m^*(t, \phi) & = 
\mathcal{F}_m^*(t) \ee^{\ii  m \phi} \eqend{.}
}
	\end{equation}
where
\begin{equation}\label{eq:def-of-Fm}
 \mathcal{F}_m(t) = \sqrt{\frac{\Gamma(m+l+1)\Gamma(m-l)}{4 \pi}} \mathsf{P}^{-m}_l(\ii \sinh t)\eqend{,}
\end{equation}
 we find
		\begin{equation}
		\eqalign{	(f_m, f_{m'}) & = - (g^*_m, g^*_{m'}) =\delta_{mm'}\eqend{,}\\
                                (f_m, g_{m'}^*) & = 0\eqend{.}
}
		\end{equation}

For the space $\mathcal{S}_\beta^+$ (and hence $\mathcal{S}_\beta^-$) to be invariant under the rotations $\phi \mapsto \phi+\alpha$ for all $\alpha\in\mathbb{R}$, the spaces $\mathcal{S}_\beta^+$ and $\mathcal{S}_\beta^-$ must be spanned by $\{F_m\}_{m\in \beta+\mathbb{Z}}$ and $\{ G^*_m\}_{m\in\beta+\mathbb{Z}}$, respectively, which are of the form
	\begin{equation}\label{GeneralFm}
	\eqalign{	 F_m(t, \phi) & = \cosh \alpha_m  f_m(t,\phi) + \ee^{\ii  \gamma_m} \sinh \alpha_m  g_m^*(t,\phi) \eqend{,} \\ 
		G_m^*(t, \phi) & = \ee^{-\ii \gamma_m} \sinh \alpha_m  f_m(t,\phi) + \cosh \alpha_m  g_m^*(t,\phi) \eqend{,} 
		}
	\end{equation}
up to constant overall phase factors for $F_m$ and $G_m^*$ (see \ref{Appendix:S-beta} for a proof). These functions satisfy
		\begin{equation}\label{eq:orthonormality-FG}
	\eqalign{	(F_m, F_{m'}) & = - (G^*_m, G^*_{m'}) =\delta_{mm'}\eqend{,}\\
                                (F_m, G_{m'}^*) & = 0\eqend{.}
}
	\end{equation}
We let $\alpha_m \geq 0$ without loss of generality.
	It will be useful later to note that
	\begin{equation}\label{eq:G-F-phi}
	    G_m(t,\phi) = F_m(t,-\phi)\eqend{,}
	\end{equation}
which follows from (\ref{GeneralFm}) and $g_m(t,\phi)=f_m(t,-\phi)$.

By the completeness of the functions $F_m$ and $G_m^*$, inherited from that of the functions $f_m$ and $g_m^*$, the quantum field $\Phi(t,\phi)$ can be expanded as
\begin{equation}\label{eq:Phi-expansion}
		\Phi(t, \phi) = \sum_{m\in\beta+\mathbb{Z}} \left[ a_m F_m(t, \phi) + b_m^\dagger G_m^* (t, \phi) \right] \eqend{,}
	\end{equation}
where $a_m$ and $b_m^\dagger$ are constant operators. As is well known~\cite{birrell1984quantum, wald1994quantum, fulling1989aspects, parker2009quantum}, with~(\ref{eq:orthonormality-FG}) the equal-time commutation relations for the quantum field $\Phi(t,\phi)$ and its time derivative are equivalent to
\begin{equation}\label{eq:a-b-commutators}
		[a_m, a^\dagger_{m'}] = [b_m, b^\dagger_{m'}]=\delta_{mm'}, \qquad \forall m,m' \in \beta + \mathbb{Z} \eqend{,}
	\end{equation}
with all other commutators vanishing.  We define the vacuum state $\ket{0}$ as a state annihilated by all $a_m$ and $b_m$, i.e.
	\begin{equation}\label{eq:def-of-vac}
		a_m \ket{0} = b_m \ket{0} = 0, \qquad \forall m \in \beta+\mathbb{Z} \eqend{.}
	\end{equation}
This vacuum state is rotationally invariant since the subspace $\mathcal{S}_\beta^+$ is.

Epstein and Moschella investigated real scalar field theory on the double cover of $\mathrm{dS}_2$ (with periodic boundary condition)~\cite{epstein2018topological}.  This theory is equivalent to that consisting of two real scalar fields with $\beta=0$ and $\beta=1/2$, respectively.  These real scalar fields must satisfy the reality conditions $G_{-m}=\ee^{\ii \delta_m}F_{m}$ with $\delta_m\in\mathbb{R}$ in (\ref{eq:Phi-expansion}) because $b_m \propto a_{-m}$.  These conditions are impossible to satisfy unless $\beta=0$ or $\beta=1/2$.  
The reality condition $G_{-m}=\ee^{\ii\delta_m}F_m$ imposes some constraints on the parameters $\alpha_m$ and $\gamma_m$ for the cases $\beta=0$ and $\beta=1/2$ as follows. 
We use the connection formula~\cite[Eq.~14.9.7]{NIST:DLMF}
	\begin{equation}\label{eq:Pconnection1}
	\fl \qquad	\frac{\sin(l - m) \pi}{\Gamma(m + l + 1)} \mathsf{P}^m_l(u) = \frac{\sin l \pi}{\Gamma(l-m+1)} \mathsf{P}^{-m}_l(u) - \frac{\sin m \pi}{\Gamma(l - m + 1)}\mathsf{P}^{-m}_l(-u)\eqend{,}
	\end{equation}
and the relation $\Gamma(x)\Gamma(1-x) = \pi/\sin\pi x$ to find
\begin{equation}\label{eq:minus-in-plus}
\eqalign{ f_{-m}(t,\phi) & = \frac{1}{\sqrt{\sin^2l\pi -\sin^2\beta\pi}}\left[ \sin m\pi f^*_m(t,\phi) -\sin l\pi g_m(t,\phi)\right] \eqend{,}\\ 
 g_{-m}(t,\phi) & = \frac{1}{\sqrt{\sin^2l\pi -\sin^2\beta\pi}}\left[ \sin m\pi g^*_m(t,\phi) -\sin l\pi f_m(t,\phi)\right]\eqend{,}
}
\end{equation}
where $f_m$ and $g_m$ are defined by (\ref{eq:def-fm}) and (\ref{eq:def-of-Fm}).  Note that 
\begin{equation}
\sin l\pi < 0\eqend{,}    
\end{equation} 
by (\ref{eq:condition-on-l}).
For $\beta=0$ we have $m\in\mathbb{Z}$
and, hence, equation~(\ref{eq:minus-in-plus}) implies $g_{-m}(t,\phi) = f_m(t,\phi)$ in this case. Then, the reality condition $G_{-m}=\ee^{\ii \delta_m} F_m$ reads
$\alpha_{-m}=\alpha_m$ and $\ee^{\ii\gamma_{-m}}=\ee^{\ii\gamma_m}$.  For $\beta=1/2$
we write down the reality condition $G_{-m}=\ee^{\ii \delta_m}F_m$ in terms of $f_m$ and $g_m^*$ by substituting (\ref{GeneralFm}) and using (\ref{eq:minus-in-plus}) with $\beta=1/2$ to express $f_{-m}^*$ and $g_{-m}$ in terms of $f_m$ and $g_m^*$.  Then, by comparing the coefficients of $f_m$ and $g_m^*$ we find
\begin{equation}
\fl \quad \left(\begin{array}{c}\!
 \cosh\alpha_m \\ \ee^{\ii \gamma_m}\sinh\alpha_m\!\end{array}\right)
= \frac{\ee^{-\ii\delta_m}}{\sqrt{\sin^2l\pi -1}}
\left(\begin{array}{cc}\! - \sin l\pi & \sin m\pi \\ \sin m \pi & - \sin l\pi\!\end{array}\right)
\left(\begin{array}{c}\! \cosh \alpha_{-m} \\ \ee^{\ii \gamma_{-m}}\sinh \alpha_{-m}\!\end{array}\right) \eqend{,}
\nonumber \\ \label{eq:reality-beta-half} 
\end{equation}
for some $\delta_m\in \mathbb{R}$, which is adjusted so that the upper component on the right-hand side is real and positive.
The inverse relation implies that $\ee^{\ii\delta_{-m}} = \ee^{-\ii\delta_m}$.  These conditions for $\beta=1/2$ and the conditions $\alpha_m=\alpha_{-m}$ and $\ee^{\ii\gamma_m} = \ee^{\ii\gamma_{-m}}$ for $\beta=0$ are equivalent to those found in \cite{epstein2018topological}.

\section{De~Sitter-invariant states}\label{sec:dSinvariance}

Recall that the mode functions $F_m$ ($G_m^*$) form the subspace $\mathcal{S}_\beta^{+}$ ($\mathcal{S}_\beta^{-}$) of the solution space $\mathcal{S}_\beta$. Since these mode functions are the coefficient functions of the annihilation (creation) operators in (\ref{eq:Phi-expansion}) and since the vacuum state $\ket{0}$ is defined by (\ref{eq:def-of-vac}), the invariance of the vacuum state under the group action $\widetilde{\mathrm{SL}}(2,\mathbb{R})$ is equivalent to the invariance of $\mathcal{S}_\beta^{\pm}$ under the same
action.  Thus, the vacuum state $\ket{0}$ is invariant under $\widetilde{\mathrm{SL}}(2,\mathbb{R})$ if and only if the linear spans of two sets of functions $\{F_m\}_{m\in \beta+\mathbb{Z}}$ and $\{G_m^*\}_{m\in \beta+\mathbb{Z}}$, which are bases of the spaces $\mathcal{S}_\beta^+$ and $\mathcal{S}_\beta^{-}$, respectively, are invariant under this group.  This condition is equivalent to 
\begin{equation}\label{eq:transf-of-Fm}
		F_m \mapsto F^\prime_m = \sum_{n\in \beta+\mathbb{Z}} U_{m n} F_{n}\eqend{,}
\end{equation}
under the action of $\widetilde{\mathrm{SL}}(2,\mathbb{R})$.  (This condition for the invariance of the space $\mathcal{S}_\beta^{+}$ also makes the orthogonal subspace $\mathcal{S}_\beta^{-}$ invariant.).

The infinite-dimensional matrix $U_{mn}$ in (\ref{eq:transf-of-Fm}) is unitary because the inner product~(\ref{eq:dS2-KG-Product}) is positive definite on $\mathcal{S}_\beta^+$ and invariant under $\widetilde{\mathrm{SL}}(2,\mathbb{R})$ and $(F_m,F_{m'}) = \delta_{mm'}$.
Hence, the functions $F_m$ form a unitary representation of the universal covering group $\widetilde{\mathrm{SL}}(2,\mathbb{R})$ of $\mathrm{SO}_0(2,1)$ with the central element $R_{2\pi}$ represented by $\ee^{2\pi\ii\beta}$ and the Casimir operator $Q$ having the eigenvalue $l(l+1)$ if the vacuum state is invariant.  With the restriction (\ref{eq:condition-on-l}) on $l$ the irreducible unitary representation formed by $F_m$ 
is either in the principal or complementary series~\cite{pukanszky1964plancherel,kitaev2017notes}, the former with $l=-1/2+\ii\lambda,\lambda\geq 0$ and the latter with $-1/2\leq l < -|\beta|$.
An important fact is that the label $m$ runs through the whole set $\beta+\mathbb{Z}$ for these representations.

Now, we shall find the conditions on $\alpha_m$ and $\gamma_m$ in (\ref{GeneralFm}) for the $\widetilde{\mathrm{SL}}(2,\mathbb{R})$ invariance of the vacuum state, i.e.\ the condition for the functions $F_m$ to transform among themselves under this group as in (\ref{eq:transf-of-Fm}).  It is sufficient to examine the transformation of $F_m$ under the infinitesimal action generated by the Killing vectors $\xi_0$, $\xi_1$ and $\xi_2$ defined by (\ref{eq:dS2-KVF}).  We combine $\xi_1$ and $\xi_2$ as
\numparts
	\begin{eqnarray}\label{eq:ladders}
		\xi_+ = \xi_1 + \ii \xi_2 = \ee^{\ii  \phi} \left( \fpartial{}{t} + \ii \tanh t \fpartial{}{\phi} \right) \eqend{,} \\
		\xi_- = \xi_1 - \ii \xi_2 = \ee^{- \ii \phi} \left( \fpartial{}{t} - \ii \tanh t \fpartial{}{\phi} \right) \eqend{.}
	\end{eqnarray}
\endnumparts
Thus, our task is to find the conditions on $\alpha_m$ and $\gamma_m$ such that $\xi_0 F_m, \xi_\pm F_m \in \mathcal{S}_\beta^+$.

Since $F_m$ and $G_m^*$ are linear combinations of $f_m$ and $g_m^*$, we first determine the action of the Killing vectors $\xi_0$ and $\xi_\pm$ on $f_m$ and $g_m^*$.
It is clear that $\xi_0 f_m(t,\phi)= \ii m f_m(t,\phi)$ and $\xi_0g_m^*(t,\phi) = \ii m g_m^*(t,\phi)$.  To find the action of $\xi_{\pm}$ on $f_m$ and $g_m^*$, we use the recurrence relations~\cite[Eq.~8.733.1]{gradshteyn2014table} obeyed by the Ferrers functions to find
	\begin{equation}
	\eqalign{	\left( \sqrt{1 - u^2} \frac{d}{d u}  - \frac{mu}{\sqrt{1 - u^2}} \right) \mathsf{P}^{-m}_l(u) & = - \mathsf{P}^{-m+1}_{l}(u) \eqend{,} \\
		\left( \sqrt{1 - u^2} \frac{d}{d u} + \frac{mu}{\sqrt{1-u^2}} \right) \mathsf{P}^{-m}_l(u) & = (l-m)(l+m+1) \mathsf{P}^{-m-1}_l(u) \eqend{.}
		}
	\end{equation}
Then, we find, using (\ref{eq:def-fm}), (\ref{eq:def-of-Fm}) and
(\ref{eq:ladders}),
\begin{equation}
\eqalign{ \xi_+\left( \begin{array}{c} f_m(t,\phi) \\ g_m^*(t,\phi)\end{array}\right) & =  
- \ii \sqrt{(m-l)(m+l+1)}\left( \begin{array}{c} f_{m+1}(t,\phi) \\ - g_{m+1}^*(t,\phi)\end{array}\right)
\eqend{,} \\
\xi_-\left( \begin{array}{c} f_m(t,\phi) \\ g_m^*(t,\phi)\end{array}\right) & =  
- \ii \sqrt{(m-l-1)(m+l)}\left( \begin{array}{c} f_{m-1}(t,\phi) \\ - g_{m-1}^*(t,\phi)\end{array}\right)
\eqend{.}
}
\end{equation}
These formulas show that the sets $\{f_m\}_{m\in \beta+\mathbb{Z}}$ and $\{ \ee^{\ii m\pi}g_m^*\}_{m\in\beta+\mathbb{Z}}$ transform in exactly the same way under the infinitesimal transformations generated by the Killing vectors $\xi_0$ and $\xi_\pm$.
Thus, $F_m$ and $G_m$ will form bases for the same unitary representation of $\widetilde{\mathrm{SL}}(2, \mathbb{R})$ if and only if $\alpha_m = \alpha$ and $\gamma_m = \pi m + \gamma$ (unless $\alpha=0$) for constants $\alpha, \gamma$ independent of $m$ in (\ref{GeneralFm}). (As we stated before, if $l\in -1/2+\ii \mathbb{R}_0^+$, then this representation is in the principal series, and if $-1/2 < l < - |\beta|$, then it is in the complementary series.)
Thus, the vacuum state $|0\rangle$ is invariant if and only if the functions $F_m$ and $G_m^*$ are chosen as follows:
	\begin{equation}\label{eq:symmetricFm}
	 \eqalign{ F_m(t, \phi) & = 
\cosh \alpha\, f_m(t,\phi)  +  \ee^{\ii  \gamma + \ii m\pi} \sinh \alpha\,g_m^*(t,\phi)\eqend{,}  \\		 
G_m^*(t, \phi) & =  \ee^{-\ii \gamma-\ii m\pi} \sinh \alpha\, f_m(t,\phi)  + \cosh \alpha\,g_m^*(t,\phi) \eqend{.}
	}
	\end{equation}
We note that the state $|0\rangle$ cannot be invariant under $\widetilde{\mathrm{SL}}(2,\mathbb{R})$ if the parameter $l$ does not satisfy (\ref{eq:condition-on-l}) (with $M^2 = -l(l+1)>0$) because there are no unitary representations of this group unless it is satisfied.
	
Let us find the condition on $\alpha$ and $\gamma$ for (\ref{eq:symmetricFm}) to be compatible with the reality condition in the periodic ($\beta=0$) and anti-periodic ($\beta=1/2$) cases with a real field $\Phi(t,\phi)$.	
For $\beta=0$, equations (\ref{eq:symmetricFm}) are compatible with the reality condition $\alpha_{-m}=\alpha_m$ and $\ee^{\ii\gamma_{-m}}=\ee^{\ii\gamma_m}$ for the field $\Phi(t,\phi)$ to be Hermitian with no further restrictions because $\alpha_m = \alpha$ for all $m\in\mathbb{Z}$ and $\ee^{\ii\gamma_{-m}} = \ee^{\ii(\gamma-m\pi)} = \ee^{\ii(\gamma+m\pi)} = \ee^{\ii\gamma_{m}}$ in this case.
The freedom in the choice of these states is the same as for the $\alpha$-vacua~\cite{allen1985vacuum, mottola1985particle} of scalar fields in higher dimensional de~Sitter spaces.  For $\beta=1/2$, if we substitute the condition $\alpha_m=\alpha$ and $\gamma_m = \gamma+m\pi$ for the invariance of the vacuum state into the condition (\ref{eq:reality-beta-half}) for the Hermiticity of the field, we find that the unique solution is 
\begin{equation}\label{eq:reality-and-invariance}
\ee^{\ii\gamma}\sinh\alpha= -\ii(\ee^{2\lambda\pi} -1)^{-1/2}\eqend{,}
\end{equation}
where we note that the parameter $l$ constrained by (\ref{eq:condition-on-l}) cannot be real in this case and we must have $l=-1/2+\ii\lambda$, $\lambda > 0$. Equation~(\ref{eq:reality-and-invariance}) agrees with the condition found in~\cite{epstein2018topological}.

\section{De~Sitter-invariant Hadamard states}\label{sec:hadamard}

Having constructed a family of de~Sitter-invariant states, next we ask whether any of these states are physically reasonable in the sense that they obey the Hadamard condition (see, e.g.,~\cite{kay1991theorems}). The Hadamard condition can be motivated by reference to the equivalence principle. On short distance scales, the background spacetime appears flat and therefore we expect that the physical states display the same short-distance singularity as the flat-space theory. In general there is a large class of Hadamard states for theories on globally hyperbolic background spacetimes such as $\mathrm{dS}_2$~\cite{wald1994quantum, fulling1981singularity}.

In this and next sections we use the conformal time coordinate $\tau$ defined by~(\ref{eq:conformal-coordinates}). We denote the functions $\mathcal{F}_m(t)$, $F_m(t,\phi)$, $f_m(t,\phi)$ and so on given in terms of the conformal time as $\mathcal{F}_m(\tau)$, $F_m(\tau,\phi)$, $f_m(\tau,\phi)$ and so on.

In two-dimensional spacetime the two-point Wightman function $\mathcal{W}(x,x')$ for a Hadamard state can be expressed in a neighbourhood of the line $x=x'$ as
	\begin{eqnarray}
	\fl\qquad	\mathcal{W}(x;x') &= &\bra{0} \Phi(x) \Phi^\dagger(x') \ket{0}\nonumber \\
	\fl\qquad	&= & - \frac{1}{4 \pi}  V(x,x') \log \left( \frac{[\mu(x,x')]^2}{2} + \ii \epsilon\, \mathrm{sign}(x^0 - x^{\prime 0}) \right) + W(x,x')\eqend{,}
		\label{eq:HadamardDef}
	\end{eqnarray}
where $\epsilon$ is infinitesimal and positive. Here, $\mu(x,x')$ denotes the spacelike geodesic distance between $x$ and $x'$, $W(x,x')$ is a smooth function, and the smooth function $V(x,x')$ is state independent and satisfies $V(x,x)=1$.

In $\mathrm{dS}_2$ the short-distance behaviour of the scalar field is determined by the high angular-momentum mode functions. Thus, we are led to discuss the large-$|m|$ behaviour of the mode functions $F_m$ and $G_m^*$.  From the Klein-Gordon equation~(\ref{eq:KG-eq-global}) in conformal coordinates $\tau$ given by $\tan\tau = \sinh t$, we have
\begin{equation}
    \left( \frac{\partial^2\ }{\partial\tau^2} - \frac{\partial^2\ }{\partial\phi^2} + \frac{M^2}{\cos^2\tau}\right)F_m(\tau,\phi) = 0\eqend{.} 
\end{equation}
For large $|m|$ the last term can be neglected and one has
\begin{equation}
    F_m(\tau,\phi) \approx (A_m \ee^{-\ii|m|\tau} + B_m \ee^{\ii |m|\tau})\ee^{\ii m\phi}\eqend{,}
\end{equation}
where $A_m$ and $B_m$ are constants.  
The two-point function for the vacuum state $|0\rangle$ defined by (\ref{eq:def-of-vac}) with the expansion (\ref{eq:Phi-expansion}) is given by
\begin{eqnarray}
\mathcal{W}(\tau,\phi;\tau',\phi') & = \langle 0|\Phi(\tau,\phi)\Phi^\dagger(\tau',\phi')|0\rangle 
\nonumber \\
& = \sum_{m\in \beta+\mathbb{Z}}F_m(\tau,\phi)F_m^*(\tau',\phi')\eqend{.}
\end{eqnarray}
We note that 
\begin{equation}\label{eq:reversed-2pt}
    \langle 0|\Phi^\dagger(\tau,\phi)\Phi(\tau',\phi')|0\rangle = \mathcal{W}(\tau,-\phi;\tau',-\phi')\eqend{,}
\end{equation}
which follows from (\ref{eq:G-F-phi}). As is well known, the mode functions $F_m$ for large $|m|$ locally resemble the positive-frequency solutions in flat space if the two-point function $\mathcal{W}(\tau,\phi;\tau',\phi')$ is Hadamard, i.e.\ of the form (\ref{eq:HadamardDef}).
Thus, for a Hadamard state we must have $F_m(\tau,\phi)\sim \ee^{-\ii|m|\tau + \ii m \phi}$ for large $|m|$.

To discuss the large $|m|$ behaviour of $F_m(\tau,\phi)$ we first discuss that of the function $\mathcal{F}_m(\tau)$ defined by (\ref{eq:def-of-Fm}) (in conformal time coordinate), which can be written using~(\ref{eq:Pdefinition})
\begin{equation}\label{eq:mathcalFm-tau}
   \fl \qquad \mathcal{F}_m(\tau) = \sqrt{\frac{\Gamma(m+l+1)\Gamma(m-l)}{\left[\Gamma(1+m)\right]^2}}\ee^{-\ii m \tau}F\left(1+l,-l;1+m; \frac{1-\ii\tan\tau }{2}\right)\eqend{.}
\end{equation}
It will be useful to find the $|m|\to \infty$ limit of this function for $m\in\mathbb{C}$ in general for later use. For $z$ purely imaginary and $|\arg c| \leq \pi - \delta$ for some $\delta>0$, the following estimate of the hypergeometric function for large $|c|$ is valid~\cite[Eq.~15.12.2]{NIST:DLMF}:
\begin{equation}\label{eq:HG-estimate}
    F\left(a,b;c;\frac{1-z}{2}\right) = 1 + O(c^{-1})\eqend{.}
\end{equation}
We also note that Stirling's formula~\cite[Eq.~5.11.7]{NIST:DLMF} implies
\begin{equation}\label{eq:stirlings}
    \Gamma(m+b) = \sqrt{2\pi}\ee^{-m}m^{m+b-\frac{1}{2}}(1+O(|m|^{-1}))\eqend{,}
\end{equation}
if $\arg m \leq \pi - \delta$ for some $\delta>0$.  From this we find
\begin{equation}\label{eq:Gamma-estimate}
    \frac{\Gamma(m+l+1)\Gamma(m-l)}{\left[\Gamma(1+m)\right]^2} = \frac{1}{m}\left(1+O(|m|^{-1})\right)\eqend{.}
\end{equation}
We use the estimates (\ref{eq:HG-estimate}) and (\ref{eq:Gamma-estimate}) to find the asymptotic behaviour of $\mathcal{F}_m(\tau)$ given by~(\ref{eq:mathcalFm-tau})
as $m \to +\infty$. (We cannot use these estimates for the limit $m\to - \infty$ because the condition that $|\arg m| \leq \pi - \delta$ for some $\delta >0$ is not satisfied in this case.) Thus, we find
\begin{equation}\label{eq:estimate-of-Fm}
    \mathcal{F}_m(\tau) = \frac{1}{\sqrt{4\pi m}}\ee^{-\ii m\tau}\left[ 1+O(m^{-1})\right]\eqend{,}
\end{equation}
for large and positive $m$.

Now, using~(\ref{eq:estimate-of-Fm}) for $f_m$ and $g_m^*$ defined by (\ref{eq:def-fm}), we find in conformal coordinates
	\begin{equation}\label{eq:fpositiveasymp}
	\eqalign{	f_m(\tau,\phi)
		& \approx \frac{1}{\sqrt{4 \pi m}} \ee^{\ii m(\phi-\tau)}\eqend{,}\\
		g_m^*(\tau,\phi)
		& \approx \frac{1}{\sqrt{4 \pi m}} \ee^{\ii m(\phi+\tau)}\eqend{,}
		}
	\end{equation}
for large and positive $m$.  Hence, the mode functions $F_m$ defined by (\ref{GeneralFm}) are approximated for large and positive $m$ as
\begin{equation}
F_m(\tau,\phi)  \approx  \frac{1}{\sqrt{4\pi m}}\ee^{\mathrm{i}m\phi}\left( \cosh\alpha_m\,\ee^{-\ii m\tau} + \ee^{\ii\gamma_m}\sinh\alpha_m\,\ee^{\ii m\tau}\right)\eqend{.}
\end{equation}
Since we must have $F_m(\tau,\phi)\sim \ee^{-\ii m \tau + \ii m\phi}$ for the vacuum state $|0\rangle$ to be Hadamard, the parameter $\alpha_m$ cannot have a nonzero limit as 
$m\to \infty$.
For a de~Sitter-invariant state, $\alpha_m = \alpha$ is $m$-independent as we have seen in the previous section. Hence, for this state to be Hadamard as well, we must have $\alpha_m = 0$ for all $m$, i.e.\ $F_m=f_m$ for all $m$. However,
by (\ref{eq:minus-in-plus}) we find for large and negative $m$
	\begin{equation}\label{eq:fnegativeasymp}
\fl \qquad	f_m(\tau, \phi) \approx \frac{\ee^{\ii  m \phi}}{\sqrt{4 \pi |m|(\sin^2 l \pi - \sin^2 \beta \pi})}\left( - \sin l \pi\,\ee^{-\ii |m|\tau} + \sin m \pi\,\ee^{\ii |m|\tau}\right)\eqend{,}
	\end{equation}
Thus, $F_m(\tau,\phi)=f_m(\tau,\phi) \sim \ee^{-\ii|m|\tau + \ii m\phi}$ for large and negative $m$ if and only if $m \in\mathbb{Z}$.  Since $m \in\beta+\mathbb{Z}$, this is the case if and only if $\beta=0$.  Therefore, only the periodic theory admits a de~Sitter-invariant Hadamard state, which is the Bunch-Davies vacuum state.

\section{The two-point function for de~Sitter-invariant vacuum states} \label{sec:explicit-2p-functions}

In this section we present closed-form expressions of the Wightman two-point functions $\mathcal{W}(x,x')$ for de~Sitter-invariant vacuum states for automorphic scalar fields.  It will be observed that these two-point functions are singular when the two points are at antipodal points except for the case with $\beta=0$ and $\alpha=0$ (the Bunch-Davies vacuum state).  Radzikowski~\cite{radzikowski1996hadamard} has proved that a Hadamard state has no non-local singularities for non-automorphic fields in globally hyperbolic spacetime.  It is not clear if his result applies to automorphic fields, but it is interesting that the de~Sitter-invariant states with a non-local singularity are not Hadamard for automorphic fields in $\mathrm{dS}_2$.

For a state invariant under $\widetilde{\mathrm{SL}}(2,\mathbb{R})$, the two-point function should be determined as a function of the geodesic distance~\cite{allen1986vector}, although this function may depend on the spatial and temporal ordering of the points because the two-point function is not necessarily invariant under the discrete de~Sitter transformations. Such a de~Sitter-invariant two-point function solves the Klein-Gordon equation
	\begin{equation}
		(\Box - M^2) \mathcal{W}(\mu) = 0\eqend{,}
	\end{equation}
which can be rewritten as the Legendre equation in $Z(x,x') = \cos \mu(x,x')$~\cite{allen1986vector} with $M^2=-l(l+1) >0$:
\begin{equation}\label{eq:legendre-Z}
    \left[(1-Z^2)\frac{\D^2\ }{\D Z^2} - 2Z\frac{\D\ }{\D Z} + l(l+1)\right]\mathcal{W} = 0\eqend{,}
\end{equation}
where $\mu(x,x')$ is the geodesic distance if $x$ and $x'$ are spacelike separated [see (\ref{eq:defZ})]. Therefore, a de~Sitter-invariant two-point function is a linear combination of $\mathsf{P}_l(-\cos\mu)$ and $\mathsf{P}_l(\cos\mu)$ 
if $\cos\mu\neq \pm 1$.
The periodicity of the fields is accounted for by extending the two-point function as
	\begin{equation}
		\mathcal{W}(\tau,\phi+2 \pi M;\tau',\phi' + 2 \pi N) = \ee^{2 \pi \ii(M-N)  \beta} \mathcal{W}(\tau,\phi;\tau',\phi'),
	\end{equation}
where $M,N \in \mathbb{Z}$.

Let us define
\begin{equation} \label{eq:W-beta-for-fm}
\mathcal{W}_\beta^{(0)}(\tau,\phi) := \sum_{m\in \beta+\mathbb{Z}}f_m(\tau,\phi)f^*_m(0,0)\eqend{.}
\end{equation}
The two-point function for the de~Sitter-invariant vacuum state with the mode functions $F_m$ and $G_m^*$ given by (\ref{eq:symmetricFm}) spanning the subspaces $\mathcal{S}_\beta^+$ and $\mathcal{S}_\beta^-$ is 
\begin{equation}
\fl \quad\eqalign{ \mathcal{W}(\tau,\phi)  & =  \langle 0|\Phi(\tau,\phi)\Phi^\dagger(0,0)|0\rangle  \\
&  =   \sum_{m\in \beta+\mathbb{Z}}\Big{\{}\cosh^2\alpha\, f_m(\tau,\phi)f_m^*(0,0) + \sinh^2\alpha g_m^*(\tau,\phi)g_m(0,0)  \\
&  \qquad\ \ \ \ \ \ \  + \cosh\alpha\sinh\alpha  \\
& \qquad\ \ \ \ \ \ \ \ \ \ \ 
\times \left[ \ee^{\ii(\gamma+m\pi)}g_m^*(\tau,\phi)f_m^*(0,0) + \ee^{-\ii(\gamma+m\pi)}f_m(\tau,\phi)g_m(0,0)\right]\Big{\}}\eqend{.}
}
\end{equation}
This function contains all information about $\langle 0|\Phi(\tau,\phi)\Phi^\dagger(\tau',\phi')|0\rangle$ for any $(\tau',\phi')$ because the state $|0\rangle$ is $\widetilde{\mathrm{SL}}(2,\mathbb{R})$ invariant.
We find from (\ref{eq:def-fm}) that $f_m(0,0)=g_m(0,0)\in\mathbb{R}$.  We also note that
$\ee^{-\ii m\pi}f_m(\tau,\phi) = f_m(\tau,\phi-\pi)$.  Furthermore, since $g_m^*(\tau,\phi) = f_m^*(\tau,-\phi)$, we find
$\ee^{\ii m\pi}g^*_m(\tau,\phi) = f_m^*(\tau,-\phi-\pi)$.  Thus,
\begin{equation}\label{eq:W-t-phi}
\fl \qquad \eqalign{
\mathcal{W}(\tau,\phi)  =  & \cosh^2\alpha\, \mathcal{W}^{(0)}_\beta(\tau,\phi) + \sinh^2\alpha \,\mathcal{W}^{(0)*}_\beta(\tau,-\phi) \\
& + \cosh\alpha\sinh\alpha\left[ \ee^{\ii\gamma}\mathcal{W}^{(0)*}_\beta(\tau,-\phi-\pi) + \ee^{-\ii\gamma}\mathcal{W}^{(0)}_\beta(\tau,\phi-\pi)\right]\eqend{.}
}
\end{equation}
Therefore, to find $\mathcal{W}(\tau,\phi)$ we only need to evaluate $\mathcal{W}_\beta^{(0)}(\tau,\phi)$.
We note that the commutator function,
\begin{equation}\label{eq:commutator-function}
\eqalign{ \langle 0|\left[ \Phi(\tau,\phi),\Phi^\dagger(0,0)\right]|0\rangle
& =   \mathcal{W}(\tau,\phi)-\mathcal{W}^*(\tau,-\phi) \\
& =  \mathcal{W}_\beta^{(0)}(\tau,\phi) - \mathcal{W}_\beta^{(0)*}(\tau,-\phi)\eqend{,}
}
\end{equation}
 is independent of the constants $\alpha$ and $\gamma$ as it should be.
[Here, we have used the relation $\langle 0|\Phi^\dagger(0,0)\Phi(\tau,\phi)|0\rangle = \langle 0|\Phi^\dagger(\tau,\phi)\Phi(0,0)|0\rangle^*$ and (\ref{eq:reversed-2pt}).]

To determine $\mathcal{W}_\beta^{(0)}(\tau,\phi)$ 
we exploit the fact that it is a linear combination of $\mathsf{P}_l(\cos\mu)$ and $\mathsf{P}_l(-\cos\mu)$, i.e.
\begin{equation}\label{eq:W-intermsof-P}
\mathcal{W}_\beta^{(0)}(\tau,\phi) = A_\beta \mathsf{P}_l(-\cos\mu) + B_\beta \mathsf{P}_l(\cos\mu)\eqend{,}
\end{equation}
if $\cos\mu\neq \pm 1$ as we stated before.
The coefficients $A_\beta$ and $B_\beta$ can be determined by examining
the logarithmic singularities of $\mathcal{W}_\beta^{(0)}(\tau,\phi)$, which can be found from its mode-sum expression.
 If $l$ is not an integer, the Legendre function $\mathsf{P}_l(x)$ is singular at $x=-1$ with a branch cut from $x = -1$ to $-\infty$, and $\mathsf{P}_l(1) = 1$. An expression for $\mathsf{P}_l(x)$ useful in examining its behaviour as $x \to -1$ is
\begin{equation} \label{eq:Pasymptotic}
\mathsf{P}_l(x) =\left[ - \frac{\sin l \pi}{\pi}\log \left( \frac{1 - x}{1+ x} \right)  + C_l\right]
\mathsf{P}_l(-x) - \frac{2}{\pi}\sin l\pi\,R_l(x)\eqend{,}
\end{equation}
which implies
\begin{equation}\label{eq:the-limit-x-minusone}
\mathsf{P}_l(x) = \frac{\sin l\pi}{\pi}\log(1+x) + O(1)\eqend{,}
\end{equation}
where
\numparts
\begin{eqnarray}
R_l(x) & = &  \lim_{m\to 0}\frac{\partial\ }{\partial m}F\left(-l,l+1;1-m; \frac{1+x}{2}\right)\eqend{,}
\label{eq:def-of-Rl}
\\
C_l &  = & \frac{2\sin l\pi}{\pi}\left[ \gamma + \psi(l+1)\right]+\cos l\pi\eqend{.}
\end{eqnarray}
\endnumparts
Here, $\gamma$ is Euler's constant and $\psi$ denotes the digamma function. Equation~(\ref{eq:Pasymptotic}) is derived in 
\ref{Appendix:formula-for-Pl}. We note that the function $R_l(x)$ is analytic at $x=-1$.

The two points are spacelike separated in the region shaded in green in Fig.~\ref{fig:CarterPenrose}, and the function
$\mathcal{W}_\beta^{(0)}(\tau,\phi)$ becomes singular as the point $(\tau,\phi)$ approaches one of the boundary lines of this region, $\phi\pm\tau = 0$ and $\pi - \phi \pm\tau = 0$.
By (\ref{eq:defZ}) we have $\cos \mu = \cos\phi\sec\tau$.  Hence,
\begin{equation}
    \fl \quad \log(1 - \cos\mu) \approx \log(\phi+\tau) + \log(\phi-\tau)\ \textrm{as} \ \phi \pm \tau \to 0\eqend{,}
\end{equation}
and
\begin{equation}
\fl \quad   \log(1+\cos\mu) \approx\left\{ \begin{array}{ll}\log(\pi - \phi + \tau)+\log(\pi -\phi -\tau) &  \textrm{as}\ \pi - \phi \pm \tau \to 0\eqend{,} \\
\log(\pi + \phi + \tau)+\log(\pi + \phi - \tau) & 
\textrm{as}\ \pi + \phi \pm \tau \to 0\eqend{,} 
\end{array}\right.
\end{equation}
Then, by (\ref{eq:the-limit-x-minusone}) we find the logarithmic singularities in $\mathcal{W}^{(0)}_\beta$ as the point $(\tau,\phi)$ approaches a boundary line as follows:
\begin{equation}\label{eq:W-logarithmic}
\fl \qquad \eqalign{
\mathcal{W}_\beta^{(0)}(\tau,\phi) \approx  &  \frac{A_\beta}{\pi}\sin l\pi \left[ \log(|\phi|+\tau) + \log(|\phi|-\tau)\right]\\
& + \frac{B_\beta}{\pi}\sin l\pi\left[\log(\pi -|\phi| + \tau) + \log(\pi-|\phi|-\tau)\right]\eqend{,}\ \ -\pi < \phi < \pi\eqend{,}
}
\end{equation}
in this region where the points $(\tau,\phi)$ and $(0,0)$ are spacelike separated.

To determine the constants $A_\beta$ and $B_\beta$ we find the the logarithmic singularities in $\mathcal{W}_\beta^{(0)}(\tau,\phi)$ using its mode expansion~(\ref{eq:W-beta-for-fm}):
\begin{equation}\label{eq:ModeSumWbeta}
\mathcal{W}_\beta^{(0)}(\tau,\phi) =  \sum^{\infty}_{n = 0} f_{m}(\tau, \phi) f^*_{m}(0, 0) + \sum_{n = 0}^\infty f_{- m'}(\tau, \phi) f^*_{-m'}(0, 0)\eqend{,} 
\end{equation}
where $m=n+\beta$ as before and $m'=n+1-\beta$. By (\ref{eq:minus-in-plus}) we find
\begin{equation}
\fl \qquad \eqalign{\mathcal{W}_\beta^{(0)}(\tau,\phi)\\
= \sum_{n=0}^\infty f_m(\tau,\phi)f_m^*(0,0) + \frac{\sin^2 l\pi}{\sin^2l\pi - \sin^2\beta\pi} \sum_{n=0}^\infty g_{m'}(\tau,\phi)g_{m'}^*(0,0) \\
\ \ \ + \frac{\sin^2\beta\pi}{\sin^2 l\pi - \sin^2\beta\pi}\sum_{n=0}^\infty f_{m'}^*(\tau,\phi)f_{m'}(0,0) \\
\ \ \  + \frac{\ee^{\pm \ii \beta\pi}\sin l\pi \sin \beta\pi}{\sin^2l\pi - \sin^2\beta\pi}
\sum_{n=0}^\infty \left[ f_{m'}^*(\tau,\phi)g_{m'}^*(0,0) + g_{m'}(\tau,\phi)f_{m'}(0,0)\right]\ee^{\pm \ii m'\pi}\eqend{,}
}
\end{equation}
where we have used $\e^{\pm \ii\beta\pi}\sin\beta\pi \ee^{\pm \ii m'\pi} = -\sin m'\pi$.  Then by (\ref{eq:fpositiveasymp}) we obtain
\begin{equation}
\fl \qquad \eqalign{4\pi\mathcal{W}_\beta^{(0)}(\tau,\phi)  \approx \sum_{n=n_0}^\infty \frac{1}{m}\ee^{\ii m(\phi-\tau)} + \frac{\sin^2 l\pi}{\sin^2l\pi - \sin^2\beta\pi} \sum_{n=n_0}^\infty \frac{1}{m'}\ee^{-\ii m'(\phi+\tau)} \\
\ \ \ \ \ \ \ \ \ \ \ \ \ \ \ \  + \frac{\sin^2\beta\pi}{\sin^2 l\pi - \sin^2\beta\pi}\sum_{n=n_0}^\infty \frac{1}{m'}\ee^{-\ii m'(\phi - \tau)} \\
\ \ \ \ \ \ \ \ \ \ \ \ \ \ \ \  + \frac{\ee^{\pm \ii \beta\pi}\sin l\pi \sin \beta\pi}{\sin^2l\pi - \sin^2\beta\pi}
\sum_{n=n_0}^\infty \frac{1}{m'}\left[
\ee^{-\ii m'(\phi\mp \pi - \tau)} + \ee^{-\ii m'(\phi\mp\pi+\tau)}\right]\eqend{,}
}
\end{equation}
where $n_0$ is any positive integer since only the large-$n$ parts of the series contribute to logarithmic singularities.  
The logarithmic singularity of the first sum can be found for $\phi-\tau\approx 0$ as follows:
	\begin{equation}
		\eqalign{ \sum^{\infty}_{n = n_0} \frac{1}{n + \beta} \ee^{- \ii(n + \beta) (\tau - \phi - \ii \epsilon)} &= - \ee^{- \ii \beta(\tau - \phi)} \left[ \log \left( 1 - \ee^{-\ii (\tau - \phi - \ii \epsilon)} \right) + \cdots \right] \\ 
		&\approx - \log(\phi - \tau + \ii \epsilon) + \cdots \eqend{,} }
	\end{equation}
where the terms omitted are finite as $\tau\to \phi$. We have inserted the convergence factor $\ee^{- (n+\beta)\epsilon}$, which determines how the analytic continuation is performed beyond the singularity.  The logarithmic singularities of the other terms can be found in a similar manner. 
Thus, we find
\begin{equation}\label{eq:W-log-singularities}
\fl \eqalign{ -4\pi\mathcal{W}_\beta^{(0)}(\tau,\phi) 
& \approx \
\log(\phi-\tau +\ii\epsilon) + \frac{\sin^2\beta\pi}{\sin^2l\pi - \sin^2\beta\pi}\log(\phi-\tau-\ii\epsilon) \\
& \quad + \frac{\sin^2l\pi}{\sin^2l\pi-\sin^2\beta\pi}\log(\phi+\tau-\ii\epsilon)\\
& \quad + \frac{\ee^{\pm \ii\beta\pi}\sin l\pi \sin\beta\pi}{\sin^2l\pi - \sin^2\beta\pi}\left[\log(\phi\mp\pi -\tau-\ii\epsilon) + \log(\phi\mp\pi+\tau-\ii\epsilon)\right]\eqend{.}
}
\end{equation}
By comparing this equation, disregarding the $\ii\epsilon$ prescription since it is irrelevant for spacelike separated points, and (\ref{eq:W-intermsof-P}) with the logarithmic singularities given by (\ref{eq:W-logarithmic}), we find for $-\pi < \phi < \pi$
\begin{equation}\label{eq:W-spacelike}
\fl	-4\mathcal{W}_\beta^{(0)}(\tau, \phi) = \frac{\sin l \pi}{\sin^2 l \pi - \sin^2 \beta \pi}  \mathsf{P}_l(-\cos \mu) + \frac{\ee^{\sigma \ii  \beta \pi} \sin \beta \pi}{\sin^2 l \pi - \sin^2 \beta \pi} \mathsf{P}_l(\cos \mu)\eqend{,}
	\end{equation}
where $\sigma = \phi/|\phi|$,
if $(\tau,\phi)$ and $(0,0)$ are spacelike separated.

To discuss this two-point function in the regions which are not connected to $(0,0)$ by spacelike geodesics, we need to find how it is analytically continued beyond the boundaries of the region where equation~(\ref{eq:W-spacelike}) is valid.
Equation~(\ref{eq:W-log-singularities}) implies that the function $\mathsf{P}_l(-\cos\mu)$ in (\ref{eq:W-spacelike}) is given near the future boundaries, i.e.\ near the line $\phi-\tau=0$ for $0<\phi<\pi$ and near the line $\phi+\tau=0$ for $-\pi < \phi< 0$, as follows:
		\begin{equation}
\fl \quad \frac{\sin l\pi\,\mathsf{P}_l(-\cos\mu)}{\sin^2l\pi - \sin^2\beta\pi} = \cases{ 
\frac{1}{\pi}\left[
\log(\phi-\tau+\ii\epsilon) + \frac{\sin^2\beta\pi}{\sin^2l\pi-\sin^2\beta\pi}
\log(\phi-\tau-\ii\epsilon)\right]\\
\qquad \times \mathsf{P}_l(\cos\mu) + \cdots, \ \ \textrm{for}\ \ 0 < \phi < \pi\eqend{,}\\
\frac{\sin^2 l \pi}{\sin^2 l \pi - \sin^2 \beta \pi}\times \frac{1}{\pi}\log(-\phi-\tau+\ii\epsilon)\mathsf{P}_l(\cos \mu)+ \cdots \\ \qquad \textrm{for}\ \ -\pi < \phi < 0\eqend{.}
		}
	\end{equation}
where the terms omitted are analytic at $\cos\mu=1$.  By using
	\begin{equation}
		\log(-x \pm \ii \epsilon) = \log |x| \pm \ii \pi\eqend{,}\ \ x < 0\eqend{,}
	\end{equation}
we have inside the future light-cone where $\tau \pm \phi>0$
		\begin{equation}\label{eq:future-interm}
\fl \quad \frac{\sin l\pi\,\mathsf{P}_l(-\cos\mu)}{\sin^2l\pi - \sin^2\beta\pi} = \cases{ 
\frac{\sin l\pi}{\sin^2l\pi-\sin^2\beta\pi}\widetilde{P}_l(-\cos\mu)\\
+ \frac{\ii (\sin^2 l\pi - 2\sin^2\beta\pi)}{\sin^2l\pi - \sin^2\beta\pi}\mathsf{P}_l(\cos\mu)\eqend{,}\  \textrm{from}\  0 <\phi < \pi\eqend{,}\\ 
\frac{\sin l \pi}{\sin^2 l \pi - \sin^2 \beta \pi}\widetilde{P}_l(-\cos\mu)\\
+\frac{\ii\sin^2 l\pi}{\sin^2l\pi - \sin^2\beta\pi}\mathsf{P}_l(\cos\mu)\eqend{,}\ \textrm{from}\ -\pi < \phi < 0\eqend{,}
		}
	\end{equation}
where, for $x>1$,
\begin{eqnarray}
    \widetilde{P}_l(-x) & := & \frac{1}{2}\left[ \mathsf{P}_l(-x+\ii\epsilon) + \mathsf{P}_l(-x - \ii\epsilon)\right]\nonumber \\
    & = & \left[ - \frac{\sin l \pi}{\pi}\log \left( \frac{x + 1}{x-1} \right)  + C_l\right]
\mathsf{P}_l(x) -\frac{2}{\pi}\sin l\pi R_l(-x)\eqend{,}
\label{eq:widetildeP}
\end{eqnarray}
[see (\ref{eq:Pasymptotic})].  We substitute (\ref{eq:future-interm}) into (\ref{eq:W-spacelike}) and find that the result is the same whether $\mathcal{W}_\beta^{(0)}$ is analytically continued from the region with $0<\phi < \pi$ or with $-\pi < \phi < 0$ as it should be. The two-point function $\mathcal{W}_\beta^{(0)}$ in the past light-cone is found similarly
and we find
\begin{equation}
		\eqalign{ -4\mathcal{W}^{(0)}_\beta(\tau, \phi) &= \frac{\sin l \pi}{\sin^2 l \pi - \sin^2 \beta \pi} \widetilde{P}_l(- \cos \mu) \\
		&\qquad  + \left[\frac{\cos \beta \pi \sin \beta \pi}{\sin^2 l \pi - \sin^2 \beta \pi} \pm \ii\right] \mathsf{P}_l(\cos \mu) \eqend{,}\ |\phi|\mp \tau < 0 \eqend{.} }
	\end{equation}
	
Finally, we determine the two-point function in the future and past light-cones of the antipodal point $(\tau, \phi) = (0, \pi)$. There are no geodesics connecting the points in these regions to the origin and the function $\cos \mu < -1$ is defined through~(\ref{eq:defZ}) as $\cos\mu=\cos\phi/\cos\tau$.
In this case, from (\ref{eq:W-log-singularities}) we find that the Legendre function $\mathsf{P}_l(\cos\mu)$ for $0< \phi < \pi$ in (\ref{eq:W-spacelike}) behaves at the boundaries $\pi-\phi\pm\tau=0$ ($\cos\mu=-1$) as
\begin{equation}
\eqalign{
    \mathsf{P}_l(\cos\mu) & = \frac{\sin l\pi}{\pi}\left[ \log(\pi -\phi+\tau+\ii\epsilon) + \log(\pi-\phi-\tau+\ii\epsilon)\right]\\
    & \qquad \times \mathsf{P}_l(-\cos\mu)+\cdots\eqend{,}
    }
\end{equation}
where the terms omitted are analytic at $\cos\mu=-1$.  Thus, the function $\mathsf{P}_l(\cos\mu)$ is analytically continued to the region with $\cos\mu < -1$ as
\begin{equation}
    \mathsf{P}_l(\cos\mu) = \widetilde{P}_l(\cos\mu) + \ii\sin l\pi\mathsf{P}_l(-\cos\mu)\eqend{,}
\end{equation}
where $\widetilde{P}_l(-x)$ with $x>1$ is defined by (\ref{eq:widetildeP}).
By substituting this equation into (\ref{eq:W-spacelike}) we find
\begin{equation}\label{eq:W-non-geodesic}
 \fl  \eqalign{ & -4\mathcal{W}_\beta^{(0)}(\tau,\phi) \\
 & = \frac{\ee^{\ii\beta\pi}}{\sin^2l\pi-\sin^2\beta\pi} \left[\sin\beta\pi\widetilde{P}_l(\cos\mu) + \cos\beta\pi\sin l\pi \mathsf{P}_l(-\cos\mu)\right]\eqend{,}\  |\tau|> |\pi-\phi|\eqend{.}
 }
\end{equation}
This two-point function can be found inside the light-cones of the point $(0,-\pi)$ in a similar manner, and we find that it is identical to (\ref{eq:W-non-geodesic}) except that the phase factor $\ee^{\ii \beta\pi}$ is changed to $\ee^{-\ii \beta\pi}$.  This result is consistent with the automorphic boundary condition $\Phi(\tau,\phi+2\pi) = \ee^{2\ii\beta\pi}\Phi(\tau,\phi)$.

The commutator function~(\ref{eq:commutator-function}) is found as
\begin{equation}    
\langle 0|\left[\Phi(\tau,\phi),\Phi^\dagger(0,0)\right]|0\rangle = \left\{ \begin{array}{ll} -\frac{\ii}{2}\textrm{sign}(\tau)\mathsf{P}_l(\cos\mu)
& \textrm{for}\ |\tau| > |\phi|\eqend{,}\\
    0 & \textrm{otherwise} \eqend{,}
    \end{array} \right.
\end{equation}
where we have used the fact that $\mathsf{P}_l(x)$ and $\widetilde{P}_l(x)$ are real if $-1 < x$ and $x < -1$, respectively.
For $\beta=0$ (the periodic case) the de~Sitter-invariant Hadamard two-point function is given by
\begin{equation}\label{eq:BD-2pt-function}
    \mathcal{W}_0^{(\mathrm{H})}(\mu) = -\frac{1}{4\sin l\pi}\mathsf{P}_l(-\cos\mu +\ii\epsilon\tau)\,,
\end{equation}
as is well known.

Next we write down the two-point function for the unique de~Sitter-invariant state for the Hermitian field with $\beta=1/2$. In this case we have $l = - 1/2+\ii \lambda$, $\lambda > 0$, by (\ref{eq:condition-on-l}).   By substituting (\ref{eq:reality-and-invariance}) into (\ref{eq:W-t-phi}) we find this two-point function as
\begin{equation}\label{eq:beta-half}
    \eqalign{ \mathcal{W}_{1/2}^{(\mathrm{R})}(\tau,\phi) & = \frac{1}{\ee^{2\lambda\pi}-1}\left[ \ee^{2\lambda\pi}\mathcal{W}_{1/2}^{(0)}(\tau,\phi)+ \mathcal{W}_{1/2}^{(0))*}(\tau,-\phi)\right]\\
    & \qquad + \frac{\ii\ee^{\lambda\pi}}{\ee^{2\lambda\pi} - 1}\left[\mathcal{W}^{(0)}_{1/2}(\tau,\phi-\pi) + \mathcal{W}^{(0)*}_{1/2}(\tau,\pi-\phi)\right]\eqend{,}
    }
\end{equation}
where we used $\mathcal{W}_{1/2}^{(0)}(\tau,-\pi-\phi) = - \mathcal{W}_{1/2}^{(0)}(\tau,\pi-\phi)$.
If $(\tau,\phi)$ is spacelike separated from $(0,0)$, then $\mathcal{W}^{(0)}_{\beta}(\tau,-\phi) = \mathcal{W}^{(0)*}_{\beta}(\tau,\phi)$ for any $\beta$ because the commutator function~(\ref{eq:commutator-function}) vanishes in this case.  Then, substituting (\ref{eq:W-spacelike}) with $\beta=1/2$ into (\ref{eq:beta-half}) we find
\begin{equation}
\mathcal{W}^{(\mathrm{R})}_{1/2}(\tau,\phi) =  \frac{1}{4\sinh\lambda\pi}\mathsf{P}_l(-\cos\mu)\eqend{,}
\end{equation}
both for $-\pi < \phi<0$ and $0 <\phi < \pi$.  In the future and past light-cones of the origin, we find
\begin{equation}
\fl \mathcal{W}_{1/2}^{(\mathrm{R})}(\tau,\phi) =  \frac{1}{4}\left[ \frac{1}{\sinh\lambda\pi}\widetilde{P}_l(-\cos\mu) 
- \ii\,\mathrm{sign}(\tau)\, \mathsf{P}_l(\cos\mu)\right]\eqend{,}\ |\tau| > |\phi|\eqend{.}
\end{equation}
Finally, we find that this two-point function vanishes in the future and past light-cones of the point $(0,\pi)$.  Thus,
$\mathcal{W}_{1/2}^{(\mathrm{R})}(\tau,\phi) = 0$ for $\cos\mu < -1$ and
\begin{equation}\label{eq:2p-for-anti-per}
\fl \qquad \mathcal{W}_{1/2}^{(\mathrm{R})}(\tau,\phi) = \frac{1}{2(\ee^{2\lambda\pi} - \ee^{-2\lambda\pi})}\left[ \ee^{\lambda\pi} \mathsf{P}_l(-\cos\mu + \ii\epsilon\tau)
+\ee^{-\lambda\pi}\mathsf{P}_l(-\cos\mu-\ii\epsilon\tau)\right]\eqend{,}
\end{equation}
for $\cos\mu > -1$.

\section{De~Sitter non-invariant Hadamard states}\label{sec:noninvhad}

As we saw in section~\ref{sec:hadamard}, it is possible to have de~Sitter-invariant Hadamard states only for the periodic field. As the Hadamard condition is a common criterion to require for physically acceptable states, we next investigate a non-invariant Hadamard state for all values of $l$ and $\beta$ satisfying~(\ref{eq:condition-on-l}). 

The argument in section~\ref{sec:hadamard} shows that the vacuum state corresponding to the mode functions
	\begin{eqnarray}
		\label{eq:HadamardF1} F_m(\tau,\phi) &= \sqrt{\frac{\Gamma(|m|-l) \Gamma(|m|+l+1)}{4 \pi}} \mathsf{P}^{-|m|}_l(\ii \tan\tau ) \ee^{\ii  m \phi}, \\
		\label{eq:HadamardG1} G_m^*(\tau,\phi) &= \sqrt{\frac{\Gamma(|m|-l) \Gamma(|m|+l+1)}{4 \pi}} \mathsf{P}^{-|m|}_l(-\ii \tan\tau ) \ee^{\ii  m \phi},
	\end{eqnarray}
has the correct behaviour as $|m| \to \infty$ for the corresponding vacuum state to be Hadamard [see the discussion after (\ref{eq:fpositiveasymp})]. In the periodic case ($\beta=0$) these are precisely the mode functions which give the Bunch-Davies vacuum. In this section we prove that the corresponding two-point function indeed has the Hadamard form~(\ref{eq:HadamardDef}) for any $\beta$ by showing that it has the same singularities as that for the Bunch-Davies vacuum state for $\beta=0$.

The mode-sum form for the two-point function for the corresponding state is
	\begin{equation}\label{eq:tildeWmodesum}
\fl	\qquad \eqalign{	& \widetilde{\mathcal{W}}_\beta(\tau,\phi; \tau', \phi')\\
&= \sum_{n = 0}^{\infty} \frac{\Gamma(m - l)\Gamma(m + l + 1)}{4 \pi} \mathsf{P}^{-m}_l(\ii \tan \tau) \mathsf{P}_l^{-m}(-\ii \tan \tau') \ee^{\ii m(\phi - \phi')} \\
&\quad + \sum_{n = 0}^\infty \frac{\Gamma(m^\prime - l)\Gamma(m^\prime + l + 1)}{4 \pi} \mathsf{P}^{-m^\prime}_l(\ii \tan \tau) \mathsf{P}_l^{-m^\prime}(-\ii \tan \tau') \ee^{-\ii m^\prime (\phi - \phi')} \eqend{,}
}
	\end{equation}
where $m = n + \beta$ and $m^\prime = n+1 - \beta$. By recalling the definition of the Ferrers function~(\ref{eq:Pdefinition}), we can rewrite the above equation as
\begin{equation}\label{eq:modesum1}
\fl \qquad \widetilde{\mathcal{W}}_\beta(\tau,\phi; \tau', \phi') =\sum_{n=0}^{\infty} f(n+\beta;\tau,\tau')z_1^{n+\beta}+\sum_{n=0}^{\infty}f(n+1-\beta;\tau,\tau')z_2^{n+1-\beta}\,,
\end{equation}
where $z_1 := \ee^{-\ii(\tau - \tau' - \phi + \phi' - \ii \epsilon)}$, $z_2 := \ee^{-\ii(\tau - \tau' + \phi - \phi' -\ii\epsilon)}$, and
	\begin{equation}\label{eq:functionf}
\fl\quad \eqalign{f(s;\tau, \tau') := \frac{\Gamma(s-l)\Gamma(s+l+1)}{4\pi[\Gamma(1 + s)]^2} \\
\qquad\ \ \ \ \ \ \ \ \ \ \times F \left( 1+l, -l; 1 + s; \frac{1- \ii \tan \tau}{2} \right) F \left( 1+l, -l; 1 + s; \frac{1 + \ii \tan \tau'}{2} \right) \eqend{.}}
	\end{equation}
With the same definitions we also write the two-point function for the periodic case as
\begin{equation}\label{eq:modesum2}
\mathcal{W}_0(\tau,\phi; \tau', \phi') =\sum_{n=1}^{\infty} f(n;\tau,\tau')z_1^{n}+\sum_{n=1}^{\infty}f(n;\tau,\tau')z_2^{n} + f(0;\tau,\tau')\eqend{.}
\end{equation}

We now proceed to show that the two-point function in (\ref{eq:modesum1}) satisfies the Hadamard condition. As we stated before, our strategy is to show that this two-point function has exactly the same light-cone singularities as the two-point function (\ref{eq:modesum2}) for the Bunch-Davies vacuum for $\beta=0$, thus concluding that it is Hadamard.
To do so we consider the difference between these two-point functions:
\begin{equation}\label{eq:modediff}
\fl \qquad \eqalign{\Delta\widetilde{\mathcal{W}}_\beta(\tau,\phi;\tau',\phi')&:=\widetilde{\mathcal{W}}_\beta(\tau,\phi; \tau', \phi')-\mathcal{W}_0(\tau,\phi; \tau', \phi')
\\
&=\sum_{n=1}^{\infty} \left[f(n+\beta;\tau,\tau')z_1^{n+\beta}- f(n;\tau,\tau')z_1^{n}\right]\\ & \quad +\sum_{n=1}^{\infty}\left[f(n+1-\beta;\tau,\tau')z_2^{n+1-\beta}-f(n;\tau,\tau')z_2^{n}\right]\\ 
& \quad + f(\beta;\tau,\tau')z_1^\beta + f(1-\beta;\tau,\tau')z_2^{1-\beta}-f(0;\tau,\tau')\eqend{.}
}
\end{equation}

In \ref{appendix-integral} it is shown that the analytic continuations of the differences of series in (\ref{eq:modediff}) are holomorphic if $z_i\neq 0$ and $|\arg z_i| < 2\pi$, $i=1,2$, provided that $|f(s;\tau,\tau')|$ grows at most polynomially as a function of $s$
for $\mathrm{Re}\,s > 0$.  In fact, equations~(\ref{eq:HG-estimate}) and (\ref{eq:Gamma-estimate}) imply that $f(s;\tau,\tau')$ tends to $0$ as $|s|\to \infty$ with $|\arg s| < \pi - \delta$ for some fixed $\delta>0$, and hence for $\mathrm{Re}\,s > 0$.  Thus, the difference of the two-point functions, $\Delta\widetilde{W}_\beta(\tau,\phi;\tau',\phi')$, is holomorphic if $|(\tau\pm\phi)-(\tau'\pm\phi')|< 2\pi$.
This result implies that the two-point functions $\widetilde{\mathcal{W}}_\beta(\tau,\phi;\tau',\phi')$ and $\mathcal{W}_0(\tau,\phi;\tau',\phi')$ have the same singularity structure on the light-cones $\phi-\tau=\phi'-\tau'$ and $\phi+\tau=\phi'+\tau'$, with no other singularities in their neighbourhood.  Thus, the state with the two-point function $\widetilde{\mathcal{W}}_\beta$ is Hadamard. 

We note that $\Delta\widetilde{\mathcal{W}}_\beta(\tau,\phi;\tau',\phi')$ is singular on the lines $|(\phi\pm\tau)-(\phi'\pm\tau')| = 2\pi$. These singularities are a manifestation of the automorphic nature of $\widetilde{\mathcal{W}}_\beta$: at $(\phi\pm\tau)-(\phi'\pm\tau')=2\pi$ where both $\widetilde{\mathcal{W}}_\beta$ and $\mathcal{W}_0$ are singular, the two-point function $\widetilde{\mathcal{W}}_\beta$ has the singularities of the periodic two-point function $\mathcal{W}_0$ times the phase factor $\ee^{ 2\ii\pi\beta}$. Therefore, the singularities do not cancel out in the difference. Thus, this difference cannot be holomorphic on these lines. For the same reason it cannot be holomorphic on the lines  $(\phi\pm\tau)-(\phi'\pm\tau')=-2\pi$. 

\section{The Gibbons-Hawking effect in a non-invariant Hadamard state} \label{sec:gibbons-hawking-non-inv}
	
In this section we demonstrate that the non-invariant Hadamard state with the two-point function~(\ref{eq:tildeWmodesum}) approximately exhibits the Gibbons-Hawking effect for all $\beta$ including the anti-periodic ($\beta=1/2$) case, clarifying the extent to which ``the Gibbons-Hawking temperature disappears'' for $\beta=1/2$~\cite{epstein2020annales}.

Any state in global de~Sitter space gives rise to a mixed state in the static patch with the metric (\ref{eq:dS2staticmetric}). As shown in \ref{appendix-GH-effect} the field operator $\Phi(x)$ is expanded in this coordinate patch as
	\begin{equation}
		\eqalign{ \Phi(T,R) = \int^\infty_0 \mathrm{d} \omega \ &\Big[ a^{(e)}_\omega F^{(e)}_\omega(T,R) + a^{(o)}_\omega F_\omega^{(o)}(T,R) \\ &\quad + b^{(e) \dagger}_\omega F^{(e) *}_\omega(T,R) + b^{(o) \dagger}_\omega F^{(o) *}_\omega(T,R) \Big]\eqend{,}}
	\end{equation}
	where $F^{(e)}(T,R)$ and $F^{(o)}(T,R)$, both proportional to $\ee^{-\ii\omega T}$ with $\omega > 0$, are even and odd in $R$, respectively.
Here, the functions $F^{(e/o)}(T,R)$ are normalised in such a way that
\begin{equation}
    [a_\omega^{(e/o)},a_{\omega'}^{(e/o)\dagger}]=[b_\omega^{(e/o)},b_{\omega'}^{(e/o)\dagger}] = \delta(\omega-\omega')\eqend{,}
\end{equation}
with all other commutators among the annihilation and creation operators vanishing.

In this section we use the global time $t$ related to the conformal time $\tau$ by $\sinh t = \tan\tau$.  We denote $\widetilde{W}_\beta(\tau,\phi;\tau',\phi')$ and $f(s;\tau,\tau')$ defined in the previous section expressed in global time $t$ as $\widetilde{W}_\beta(t,\phi;t',\phi')$ and $f(s;t,t')$, respectively. 

For the de~Sitter non-invariant Hadamard state with the two-point function $\widetilde{\mathcal{W}}_\beta$ given by~(\ref{eq:tildeWmodesum}) we consider
\begin{equation}
\int_{-\infty}^{\infty}\int_{-\infty}^{\infty}\frac{\Dt\Dt'}{2\pi}\ee^{-\ii\omega t}\widetilde{\mathcal{W}}_\beta(t,0;t',0)\ee^{\ii \omega't'}=A(\omega,\omega')\langle b_\omega^{(e)\dagger} b_{\omega'}^{(e)}\rangle_\beta\,,
\end{equation} 
where $A(\omega,\omega')=2 \pi F_{\omega}^{(e)*}(0,0) F_{\omega^\prime}^{(e)}(0,0)$.  Here, $\langle b_\omega^{(e)\dagger}b_{\omega'}^{(e)}\rangle_\beta$ is the expectation value of the number operator $b_\omega^{(e)\dagger}b_{\omega'}^{(e)}$ in the reduced state in the static patch obtained from the de~Sitter non-invariant global Hadamard state with the two-point function $\widetilde{\mathcal{W}}_\beta$.
As described in \ref{appendix-GH-effect}, the de~Sitter-invariant Hadamard state for $\beta=0$ exhibits the Gibbons-Hawking effect~\cite{gibbons1977cosmological}.  That is,
\begin{equation}
\int_{-\infty}^{\infty}\int_{-\infty}^{\infty}\frac{\Dt\Dt'}{2\pi}\ee^{-\ii\omega t}\mathcal{W}_0(t,0;t',0)\ee^{\ii \omega't'}=A(\omega,\omega')\langle b_\omega^{(e)\dagger} b_{\omega'}^{(e)}\rangle_0\eqend{,}
\end{equation} 
where
\begin{equation}
    \langle b_\omega^{(e)\dagger} b_{\omega'}^{(e)}\rangle_0 = \frac{1}{\ee^{2\pi\omega}-1}\delta(\omega-\omega')\eqend{.}
\end{equation}

Now, suppose that the integral
\begin{equation}\label{eq:W-tilde-finite}
    G(\omega,\omega')  =\int_{-\infty}^{\infty}\int_{-\infty}^{\infty}\frac{\Dt\Dt'}{2\pi}\ee^{-\ii\omega t}\Delta\widetilde{\mathcal{W}}_\beta(t,0;t',0)\ee^{\ii \omega't'}\eqend{,}
\end{equation}
is bounded for all positive $\omega$ and $\omega'$, where
    $\Delta\widetilde{\mathcal{W}}_\beta = \widetilde{\mathcal{W}}_\beta - \mathcal{W}_0$ [see (\ref{eq:modediff})].  Then
\begin{equation}\label{eq:approximate-GH}
\langle b^{(e)\dagger}_\omega b^{(e)}_{\omega'}\rangle_\beta = \frac{1}{\ee^{2\pi\omega}-1}\delta(\omega-\omega') + \frac{G(\omega,\omega')}{A(\omega,\omega')}\eqend{.}
\end{equation}
This equation implies that the non-invariant Hadamard state with the two-point function $\widetilde{\mathcal{W}}_\beta$ exhibits the Gibbons-Hawking effect approximately in the following sense. For any compactly supported function $f(\omega)$ centred at $0$ satisfying
\begin{equation}
    \int_{-\infty}^\infty |f(\omega)|^2\D\omega = 1\eqend{,}
\end{equation}
we define the smeared annihilation operator for $\epsilon >0$ 
by 
\begin{equation}
    B^{(e)}_{\epsilon}(\omega_0):= \epsilon^{-1/2}\int_0^\infty f((\omega-\omega_0)/\epsilon)b^{(e)}_\omega\,\D\omega\eqend{.}
\end{equation}
Then
\begin{equation}
    \left[B^{(e)}_\epsilon(\omega_0),B^{(e)\dagger}_\epsilon(\omega_0)\right]=1\eqend{.}
\end{equation}
The operator $B^{(e)}_\epsilon$ annihilates and the operator $B^{(e)\dagger}_\epsilon$ creates a one-particle state of definite frequency $\omega_0$ in the limit $\epsilon \to 0$.  Equation~(\ref{eq:approximate-GH}) implies
\begin{equation}\label{eq:quasi-GibbHawk}
\langle B^{(e)\dagger}_\epsilon(\omega_0)B^{(e)}_\epsilon(\omega_0)\rangle_\beta = \frac{1}{\ee^{2\pi\omega_0}-1}+ O(\epsilon)\eqend{,}\ \textrm{for}\ \epsilon \ll 1.
\end{equation}
Thus, although the state with the two-point function $\widetilde{W}_\beta$ is not exactly thermal when restricted to the static patch, it is approximately thermal with the Gibbons-Hawking temperature $1/(2\pi)$.

Now, let us show that the integral~(\ref{eq:W-tilde-finite}) is indeed bounded and, hence, that equation~(\ref{eq:quasi-GibbHawk}) holds. We first combine
the expression~(\ref{eq:modediff}) for $\Delta\widetilde{W}_\beta$ and the integral representation proved in \ref{appendix-integral} to find
\begin{equation}\label{eq:we-want-this-integral}
\eqalign{
    \Delta\widetilde{W}_\beta(t,0;t',0) & = \ii\sin\pi\beta \int_{C_\gamma}\Ds\,\frac{f(s;t,t')z^s}{\cos\pi\beta - \cos\pi(2s-\beta)} \\
& \quad    - \ii\sin\pi\beta\int_{C_\gamma}\Ds\,\frac{f(s;t,t')z^s}{\cos\pi\beta - \cos\pi(2s+\beta)}\\
    & \quad + f(\beta;t,t')z^\beta- f(0;t,t')\eqend{,}
    }
\end{equation}
where $f(s;t,t')$ is given by (\ref{eq:functionf}) with $\sinh t = \tan\tau$ and $\sinh t'=\tan\tau'$, and $z=\ee^{-\ii(\tau-\tau')}$. The contour $C_\gamma$ is the straight line from $\gamma-\ii\infty$ to $\gamma+\ii\infty$ with $0 <\gamma < \textrm{min}(1-\beta,1+\beta)$.  Next we note that
\begin{equation}
    f(s;t,t')z^s = \frac{\Gamma(s-l)\Gamma(s+l+1)}{4\pi}\mathsf{P}_l^{-s}(\ii\sinh t)\mathsf{P}_l^{-s}(-\ii\sinh t')\eqend{,}
\end{equation}
and
\begin{equation}
\eqalign{ \left| \int\Dt\int\Dt'\ee^{-\ii\omega(t-t')}f(s;t,t')z^s\right| =  & \ \frac{|\Gamma(s-l)\Gamma(s+l+1)|}{4\pi}\\ 
& \times \left| \int dt\,\mathsf{P}_l^{-s}(\ii\sinh t)\ee^{-\ii\omega t}\right|^2\eqend{.}
}
\end{equation}
Then, the bound established in \ref{appendix:bound-Ferrers} [see (\ref{eq:bound-to-be-found})] implies
\begin{equation}\label{eq:main-bound}
\eqalign{    & \left| \int\Dt\int\Dt'\ee^{-\ii\omega(t-t')}f(s;t,t')z^s\right| \\
& \leq \frac{1}{4\pi}\left[C(\gamma)\right]^2 \ee^{-\pi\omega}\left[\cosh(\pi u/2)\right]^2\left| \frac{\left[\Gamma(s+\frac{1}{2})\right]^2}{\Gamma(l+s+1)\Gamma(s-l)}\right|\eqend{,}
}
\end{equation}
where $s=\gamma+\ii u$, $\gamma \geq 0$, $u\in\mathbb{R}$, and $C(\gamma)$ is a positive constant independent of $\omega$ and $u$.
In the integral~(\ref{eq:W-tilde-finite}) the contribution from the last two terms of (\ref{eq:we-want-this-integral}) is finite by this bound.  This bound also implies that the contribution to (\ref{eq:W-tilde-finite}) of the first two terms of (\ref{eq:we-want-this-integral}) given as integrals along the contour $C_\gamma$ is also finite.  This is because the double Fourier transform of $f(s;t,t')z^s$ grows like $\ee^{\pi u}$ for large $u=\mathrm{Im}\,s$ by (\ref{eq:main-bound}), but the denominators $\cos\pi\beta -\cos\pi(2s\mp \beta)$ in (\ref{eq:we-want-this-integral}) grow like $\ee^{2\pi u}$, thus making the integral over $s$ along $C_\gamma$ from $\gamma-\ii\infty$ to $\gamma+\ii\infty$ converge exponentially fast. [Note that the ratio of the $\Gamma$-functions in (\ref{eq:main-bound}) tends to $1$ as $u\to \pm \infty$ by (\ref{eq:stirlings}).]  Thus, the function $G(\omega,\omega')$ defined by (\ref{eq:W-tilde-finite}) is indeed bounded for all positive $\omega$ and $\omega'$. Hence, equation~(\ref{eq:quasi-GibbHawk}) holds and the non-invariant Hadamard state with the two-point function $\widetilde{\mathcal{W}}_\beta$ approximately exhibits the Gibbons-Hawking effect.

\section{Summary and discussion}\label{sec:discussion}

In this paper we studied the non-interacting automorphic scalar field in two-dimensional de~Sitter space, extending the work of Epstein and Moschella~\cite{epstein2018topological,epstein2020annales}, who studied the anti-periodic real scalar field in this spacetime.  We found that there are no states which are both de~Sitter invariant and Hadamard except in the periodic case, for which the unique state with both of these properties is the standard Bunch-Davies state. 

We constructed the two-point Wightman functions explicitly for de~Sitter-invariant non-Hadamard states for the non-periodic cases.  We found that these two-point functions are singular if the two points are antipodal from each other.  Interestingly, the unique de~Sitter-invariant state for the \textit{anti-periodic real} scalar field has a vanishing Wightman two-point function when the two points cannot be connected by a geodesic [see (\ref{eq:2p-for-anti-per})]. (In this case, the antipodal singularity is a discontinuity rather than a divergence.)  We constructed these two-point functions using the fact that the de~Sitter invariance restricts them to be  linear combinations of two known functions.

We also studied a de~Sitter non-invariant Hadamard state for each periodicity variable $\beta$, including the anti-periodic case, and showed that this state approximately exhibits the Gibbons-Hawking effect.  Thus, for automorphic scalar fields in two-dimensional de~Sitter space there are states for which the physics inside a static region does not appear very different from that in the Bunch-Davies vacuum state.  Nevertheless, it is interesting that the incompatibility of the Hadamard condition with de~Sitter invariance for the anti-periodic scalar field generalises to more general automorphic boundary conditions.  It will be interesting to find the response of the Unruh-DeWitt detector~\cite{Unruh:1976db,DeWitt:1980hx} as a function of time~\cite{Louko:2006zv,Louko:2007mu} for these states and confirm that the deviation from the thermal response is only temporary.

\ack

We thank Ugo Moschella for helpful correspondence and Jorma Louko for useful discussions. D.\ S.\ B. was supported in part by an Overseas Research Scholarship from the University of York. L.\ S.\ was supported in part by a studentship from the Engineering and Physical Sciences Research Council (EPSRC) (EP/N509802/1) and the Doctoral Prize from the Department of Mathematics, University of York.

\appendix

\section{The condition on $\mathcal{S}_\beta^{\pm}$ from rotational invariance}\label{Appendix:S-beta}

Suppose that a function $F(t,\phi)$ is in $\mathcal{S}_\beta^+$, which is assumed to be invariant under $\phi\mapsto \phi+\alpha$ for all $\alpha\in \mathbb{R}$. Let
\begin{equation}
    F(t,\phi) = \sum_{n\in \beta+\mathbb{Z}} c_n(t) \ee^{\ii n \phi}\eqend{.}
\end{equation}
Then
\begin{equation}
    c_m(t)\ee^{\ii m \phi} = \frac{1}{2\pi}\int_0^{2\pi}\ee^{-\ii m\alpha}F(t,\phi+\alpha)\mathrm{d}\alpha\eqend{,}
\end{equation}
where $m\in \beta+\mathbb{Z}$.  Since the solutions $F(t,\phi+\alpha)$ must be in $\mathcal{S}_\beta^+$ by rotational invariance of this subspace, the solution $c_m(t)\ee^{\ii m\phi}$ must be in $\mathcal{S}_\beta^+$ for all $m\in\beta+\mathbb{Z}$.  Thus, any function $F\in \mathcal{S}_\beta^+$ is a linear combination of solutions of the form $c_m(t)\ee^{\ii m\phi}$, which are themselves in $\mathcal{S}_\beta^+$.  Hence a  basis for $\mathcal{S}_\beta^+$ can be chosen to be of the form $\{ F_m\}_{m\in\beta+\mathbb{Z}}$ where the solution $F_m$ for each $m\in\beta+\mathbb{Z}$ is a linear combination of $f_m$ and $g_m^*$ defined by (\ref{eq:def-fm}).
By requiring $(F_m,F_m)=1$ without loss of generality,
we find that the solution $F_m$ must be of the form given in (\ref{GeneralFm}).
Then, by the requirements $(F_m,G_m^*) =0$ and $(G_m^*,G_m^*)=-1$, we find that the solution $G_m^*\in \mathcal{S}_\beta^-$ must be of the form given there.

\section{Derivation of (\ref{eq:Pasymptotic})}\label{Appendix:formula-for-Pl}

Two independent solutions to the Legendre equation,
\begin{equation}
\left[ (1-x^2)\frac{\D^2\ }{\D x^2} - 2x \frac{\D\ }{\D x} - l(l+1)\right]f(x) = 0\eqend{,}
\end{equation}
are $f(x)=\mathsf{P}_l(x)$ and $f(x)=\mathsf{Q}_l(x)$ for $-1 < x < 1$.  We start from the following formula~\cite[Eq.~14.9.10]{NIST:DLMF}:
\begin{equation}\label{eq:Pl-by-Ql}
\mathsf{P}_l(-x)  =  - \frac{2}{\pi}\sin l\pi \mathsf{Q}_l(x) + \cos l\pi \mathsf{P}_l(x)\eqend{.}
\end{equation}
Note that
\begin{equation}
\left[ (1-x^2)\frac{\D^2\ }{\D x^2} - 2x \frac{\D\ }{\D x} - \frac{m^2}{1-x^2} + l(l+1)\right] 
\Gamma(1-m) \mathsf{P}_l^m(x)= 0\eqend{.} \label{ass-Legendre}
\end{equation}
We differentiate (\ref{ass-Legendre}), after substituting the definition~(\ref{eq:Pdefinition}) of the Ferrers function in terms of Gauss's hypergeometric function, with respect to $m$ and take the limit $m\to 0$.  Thus, we find
\begin{equation}
\left[ (1-x^2)\frac{\D^2\ }{\D x^2} - 2x \frac{\D\ }{\D x}  + l(l+1)\right]f_l(x) = 0\eqend{,} 
\end{equation}
where
\begin{equation}
f_l(x)  =  \frac{1}{2}\mathsf{P}_l(x) \log \frac{1+x}{1-x} + R_l(-x)\eqend{,}
\end{equation}
with $R_l(x)$ defined by (\ref{eq:def-of-Rl}).  This function is analytic at $x=-1$ and has the value $R_l(-1)=0$.
Since $\mathsf{P}_l(1)=1$
and~\cite[Eq.~14.8.3]{NIST:DLMF}
\begin{equation}
\mathsf{Q}_l(x)  =  \frac{1}{2}\log\left( \frac{2}{1-x}\right) - \gamma - \psi(l+1) + O(1-x)\eqend{,}
\end{equation}
we find that $f_l(x)-\mathsf{Q}_l(x)$ is finite at $x=1$, and, hence, must be proportional to $\mathsf{P}_l(x)$.
This implies
\begin{equation}
\mathsf{Q}_l(x) = \mathsf{P}_l(x)\left[ \frac{1}{2}\log\frac{1+x}{1-x}
- \gamma - \psi(l+1)\right] + R_l(-x)\eqend{.}
\end{equation}
By substituting this formula into (\ref{eq:Pl-by-Ql}) we find (\ref{eq:Pasymptotic}) with $x$ replaced by $-x$.

\section{The integral representation of the difference of two series}\label{appendix-integral}

This appendix concerns the analyticity of the series
\begin{equation}\label{eq:diff-series}
    F(z) := \sum_{n=1}^\infty \left[f(n+\beta)z^{n+\beta} - f(n)z^n\right]\eqend{,}\ -\frac{1}{2} < \beta \leq \frac{1}{2}\eqend{,}
\end{equation}
where $f(t)$ is a function holomorphic on the half-plane $\{t\in\mathbb{C}: \mathrm{Re}\,t > 0\}$ which grows at most polynomially as $|t|\to \infty$. We show that the analytic continuation of the function $F(z)$ is holomorphic in the double Riemann sheet with $z\neq 0$ and $|\arg\,z| < 2\pi$.

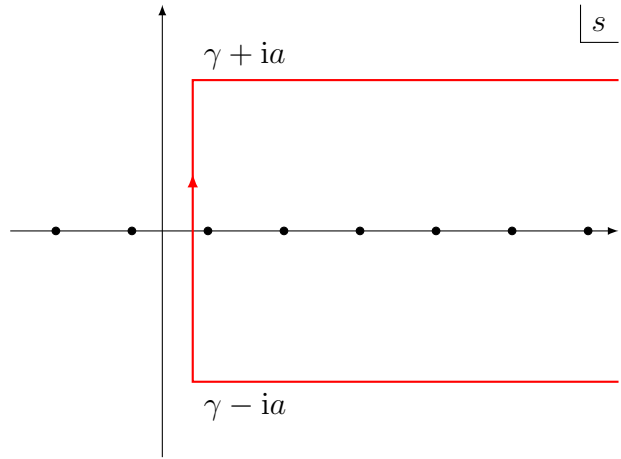
\begin{figure}[t]
		\begin{center}
			\begin{tikzpicture}
				\draw[->] (-2,0) -- (6,0);
				\draw[->] (0,-3) -- (0,3);
				
				\draw (0.4,2) node[above right] {$\gamma + \ii a$};
				\draw (0.4,-2) node[below right] {$\gamma - \ii a$};
								
				\foreach \x in {-2,...,5}
					\filldraw (\x + 0.6,0) circle (0.05);
				
				\draw[red, thick, ->- = 0.55] (6,-2) -- (0.4,-2) -- (0.4,2) -- (6,2);
				
				\draw (5.5,3) -- (5.5,2.5) node[above right] {$s$} -- (6,2.5);
			\end{tikzpicture}
		\end{center}\caption{Contour of integration $C$ in the complex $s$-plane. The black dots represent the poles of the integrand located at $s=n+\beta$, $n\in\mathbb{Z}$. \label{fig:contour}}
		\end{figure}

We start by proving an integral representation for the series
	\begin{equation}\label{S}
		S_\beta(z) = \sum^{\infty}_{n = 1} f(n + \beta) z^{n + \beta} \eqend{,}\ \ |z|<1\eqend{.}
	\end{equation}
We choose a real constant $\gamma$ satisfying $0 < \gamma < 1+\beta$.  Then, we show that this series is represented by the following contour integral:
	\begin{equation}\label{I}
		S_\beta(z) = \frac{\ii}{2} \int_C \mathrm{d} s\,f(s) z^s \cot \pi (s - \beta) \eqend{,}
	\end{equation}
where the contour $C$ consists of the straight lines connecting $\infty - \ii a$, $\gamma -\ii a$, $\gamma + \ii a$ and $\infty + \ii a$, with $a>0$, as shown in Fig.~\ref{fig:contour}.
To prove (\ref{I}) we first consider the integral $S_\beta^N(z)$ defined by replacing $C$ in (\ref{I}) by $C_N$, $N\in\mathbb{N}$, which is
the rectangular contour with vertices at $N+\gamma-\ii a$, $\gamma-\ii a$, $\gamma+\ii a$ and $N+\gamma+\ii a$. We apply the residue theorem to $S^N_\beta(z)$. We pick up the residues from the simple poles at $1 + \beta, \ldots, N-1 + \beta$ and find
	\begin{equation}\label{IN2}
		S^N_\beta(z) = \sum_{n = 1}^{N-1} f(n + \beta) z^{n+\beta} \eqend{.}
	\end{equation}
Hence, $S^N_\beta(z) \to S_\beta(z)$ as $N \to \infty$. The contribution to the integral~(\ref{IN2}) from the straight line segment from $N+\gamma+\ii a$ to $N+\gamma-\ii a$ tends to zero because $\cot\pi(s-\beta)$ on this line segment is independent of $N$ and because for $s=N+\gamma+\ii t$, $t\in [-a,a]$, we have
$|f(s)z^s| = |f(s)|\ee^{(N+\gamma)\log |z| - (\arg z)t}$. (Recall our assumptions that $|z|<1$ and that $|f(s)|$ grows at most polynomially as a function of $|s|$.)  Hence, $S_\beta^N(z)$ tends to the right-hand side of (\ref{I}).  This proves (\ref{I}).

An integral representation of the right-hand side of (\ref{eq:diff-series}) can be found from (\ref{I}) as
\begin{eqnarray}
    S_\beta(z)-S_0(z) & = &  \ii \sin\pi\beta\int_C \D s\, \frac{f(s)z^s}{\cos\pi \beta - \cos\pi(2s-\beta)}\eqend{,} \label{eq:difference-integral-rep}
\end{eqnarray}
where $0 < \gamma < \mathrm{min}(1,1+\beta)$.
Next we deform the contour $C$ by replacing the two half-lines from $\gamma\pm \ii a$ to $\infty\pm \ii a$ by the half-lines $\gamma\pm \ii a$ to $\gamma \pm \ii \infty$.  The resulting contour is the straight line parallel to the imaginary axis with $\mathrm{Re}\,s = \gamma$, that is, the straight line from $\gamma-\ii\infty$ to $\gamma+\ii \infty$, which we denote by $C_\gamma$.
This deformation is justified if the integrand decays exponentially as $|s| \to \infty$ if $\mathrm{Im}\,s \geq a$ and $\mathrm{Re}\,s \geq \gamma$ since it is holomorphic in the half-plane $\mathrm{Re}\,s > 0$ except at the poles on the real axis. 

We now show this behaviour of the integrand of (\ref{eq:difference-integral-rep}):
\begin{equation}
    G(s) : = \frac{f(s)z^s}{\cos\pi\beta - \cos\pi(2s-\beta)}\eqend{.}
\end{equation}
We first find
\begin{equation}
    |f(s)z^s| = |f(s)|\ee^{(\log|z|)\mathrm{Re}\,s - (\arg z)\mathrm{Im}\,s} \leq
    |f(s)|\ee^{(\log|z|)\mathrm{Re}\,s + (\arg z)|\mathrm{Im}\,s|}
    \eqend{.}
\end{equation}
Next, since $|\cos\pi(2s-\beta)| \approx \ee^{2\pi|\mathrm{Im}\,s|}/2$ for large
$|\mathrm{Im}\,s|$,
by choosing $a$ large enough we have for $|\mathrm{Im}\,s| \geq a$
\begin{equation}
    |\cos\pi\beta - \cos\pi(2s-\beta)| \geq \frac{1}{3}\ee^{2\pi\,|\mathrm{Im}\,s|}\eqend{,}
\end{equation}
Hence
\begin{equation}\label{eq:integral-estimate}
    |G(s)| \leq 3|f(s)|\ee^{(\log|z|)\mathrm{Re}\,s - \left[2\pi - (\arg z)\right]|\mathrm{Im}\,s|}\eqend{.}
\end{equation}
Thus, if $|z| <1$ and $|\arg z|< 2\pi$, and if $\mathrm{Re}\,s \geq \gamma$ and $|\mathrm{Im}\,s| \geq a$, then $G(s)\to 0$ exponentially as $|s|\to\infty$.  Therefore, the contour $C$ in (\ref{eq:difference-integral-rep}) can be deformed to $C_\gamma$, the straight line from $\gamma - \ii \infty$ to $\gamma + \ii \infty$, where $\gamma$ satisfies $0 < \gamma < \mathrm{min}(1,1+\beta)$.  Hence, equation~(\ref{eq:difference-integral-rep}) holds with the contour $C$ replaced by $C_\gamma$.
This integral representation
with contour $C_\gamma$ gives the analytic continuation of the function $F(z)= S_\beta(z)-S_0(z)$ from $0 < |z| < 1$ to all nonzero $z$ satisfying $|\arg z| < 2\pi$ because the estimate~(\ref{eq:integral-estimate}) shows that this integral and its derivative are convergent.

\section{The Gibbons-Hawking effect in two dimensions}\label{appendix-GH-effect}
	
In this appendix we review the Gibbons-Hawking effect in two dimensions~\cite{gibbons1977cosmological,figari1975interacting}, i.e.\ the fact that the Bunch-Davies vacuum state is a thermal state with respect to the energy defined in the static patch of de~Sitter space. The static coordinates $(T,R)$ are defined by (\ref{eq:dS2static}) and the metric in the static patch covered by these coordinates is given by (\ref{eq:dS2staticmetric}).

The Klein-Gordon equation obeyed by a scalar field $\Phi(T,R)$ with mass $M$ takes the form
	\begin{equation}\label{eq:KG-equation-dS}
		\left[ - \frac{1}{1 - R^2} \frac{\partial^2}{\partial T^2} + \fpartial{}{R} \left( 1 - R^2 \right) \fpartial{}{R} 
		- M^2 \right] \Phi(T,R) = 0\eqend{.}
\end{equation}	 
Two linearly independent solutions to this equation proportional to $\ee^{-\ii\omega T}$ are
	\begin{eqnarray}
\fl		F^{(e)}_\omega(T,R) = A^{(e)}_\omega (1 - R^2)^{\frac{\ii \omega}{2}} F \left(\frac{1 + \ii \omega + l}{2}, \frac{\ii \omega - l}{2}; \frac{1}{2}; R^2 \right) \ee^{-\ii \omega T}\eqend{,}\label{eq:staticmodes1} \\
\fl		F^{(o)}_\omega(T,R) = A^{(o)}_\omega (1 - R^2)^{\frac{\ii \omega}{2}} R F \left(\frac{2+\ii \omega + l}{2}, \frac{1 + \ii \omega - l}{2}; \frac{3}{2}; R^2 \right) \ee^{- \ii \omega T}\eqend{,}\label{eq:staticmodes2}
	\end{eqnarray}
with well-defined parity.  The normalisation constants, $A_\omega^{(e)}$ and $A_\omega^{(o)}$ are determined from the Klein-Gordon inner product
		\begin{equation}
		(\Psi_1, \Psi_2) = \ii \int_{-1}^1 \frac{\mathrm{d} R}{1 - R^2} \left( \Psi_1^* \fpartial{\Psi_2}{T} - \fpartial{\Psi_1^*}{T} \Psi_2 \right)\eqend{.}	
\end{equation}
The inner product of two solutions to (\ref{eq:KG-equation-dS})  of the form $F_\omega(R)\ee^{-\ii\omega T}$ and $\widetilde{F}_{\omega'}(R)\ee^{-\ii\omega'T}$ can be found in a manner similar to the four-dimensional case~\cite{higuchi1987gibbons-hawking} as
\begin{equation}\label{eq:inner-prod-approx}
 \fl   (F_\omega,\widetilde{F}_{\omega'}) = \lim_{\epsilon\to 0}\frac{1}{\omega-\omega'}\left[(1-R^2)\left( F_\omega(R)\widetilde{F}_{\omega'}'(R) - F'_\omega(R)\widetilde{F}_{\omega'}(R)\right)\right]_{R=-1+\epsilon}^{R=1-\epsilon}\eqend{.}
\end{equation}
We find for $R^2\approx 1$
\begin{equation}\label{eq:R1-approx}
F^{(e/o)}_\omega(T,R) \approx 2A_\omega^{(e/o)}B_\omega^{(e/o)}\cos\left[\frac{\omega}{2}\log(1-R^2) - \delta_\omega^{(e/o)}\right]\ee^{-\ii\omega T}\eqend{,}
\end{equation}
where
\begin{eqnarray}
    B_\omega^{(e)}\ee^{\ii\delta_\omega^{(e)}} & = & \frac{\sqrt{\pi}\Gamma(\ii\omega)}{\Gamma\left(\frac{1+\ii\omega+l}{2}\right)\Gamma\left(\frac{\ii\omega-l}{2}\right)}\eqend{,}\\
    B_\omega^{(o)}\ee^{\ii\delta_\omega^{(o)}} & = & \frac{\sqrt{\pi}\Gamma(\ii\omega)}{2\Gamma\left(\frac{2+\ii\omega+l}{2}\right)\Gamma\left(\frac{1+\ii\omega-l}{2}\right)}\eqend{,}
\end{eqnarray}
with $B^{(e)}_\omega$ and $B^{(o)}_\omega$ real and positive. We substitute (\ref{eq:R1-approx}) into (\ref{eq:inner-prod-approx}) and use the identity
\begin{equation}
    \lim_{\epsilon\to 0} \frac{\sin\left[(\omega-\omega')\log(2\epsilon)\right]}{\omega-\omega'} = -\pi \delta(\omega-\omega')\eqend{,}
\end{equation}
and neglect rapidly oscillating bounded functions of $\omega$ and $\omega'$ to find that for
\begin{equation}\label{eq:normalisation-cond}
    (F_{\omega}^{(e/o)},F_{\omega'}^{(e/o)}) = \delta(\omega-\omega')\eqend{,}
\end{equation}
we must have
\begin{equation}
|A^{(e/o)}_\omega|^2 = \frac{1}{8\pi\omega|B^{(e/o)}_\omega|^2}\eqend{.}
\end{equation}
That is,
\begin{eqnarray}
		A^{(e)}_\omega & = \sqrt{\frac{\sinh \pi \omega}{8 \pi^3}} \left\vert \Gamma\left(\frac{1+\ii \omega + l}{2}\right) \Gamma\left(\frac{\ii \omega - l}{2}\right) \right\vert\eqend{,}\label{eq:coeff-A-even}\\
		A^{(o)}_\omega & = \sqrt{\frac{\sinh \pi \omega}{2 \pi^3}} \left\vert \Gamma \left( \frac{2+\ii \omega + l}{2} \right) \Gamma \left(\frac{1+\ii \omega - l}{2} \right) \right\vert\eqend{,}
\end{eqnarray}
where we have used $|\Gamma(\ii\omega)|^2 = \pi/(\omega\sinh\pi\omega)$.
With this normalisation, $F^{(e)}_\omega$ and $F^{(o)}_\omega$ form an orthonormal basis for the space of solutions which are positive-frequency with respect to the time coordinate $T$ in the static patch.

In terms of these mode functions, the field operator $\Phi$ can be expanded in the static patch as
	\begin{equation}
		\eqalign{ \Phi(T,R) = \int^\infty_0 \mathrm{d} \omega \ &\Big[ a^{(e)}_\omega F^{(e)}_\omega(T,R) + a^{(o)}_\omega F_\omega^{(o)}(T,R) \\ &\quad + b^{(e) \dagger}_\omega F^{(e) *}_\omega(T,R) + b^{(o) \dagger}_\omega F^{(o) *}_\omega(T,R) \Big]\eqend{.}}
	\end{equation}
The normalisation condition (\ref{eq:normalisation-cond}) implies that the annihilation operators, $a^{(e/o)}_\omega$ and $b^{(e/o)}_\omega$, and the creation operators, $a^{(e/o)\dagger}_\omega$ and $b^{(e/o)\dagger}_\omega$, satisfy
	\begin{equation}
		[a^{(e/o)}_\omega, a^{(e/o)\dagger}_{\omega^\prime}] 
		=[b^{(e/o)}_\omega, b^{(e/o)\dagger}_{\omega^\prime}] =\delta(\omega - \omega^\prime)\eqend{,}
	\end{equation}
with all other commutators vanishing.

Now let us consider the following two-point function:
	\begin{equation}
		\mathcal{W}_0(T,R; T^\prime, R^\prime) = \langle \Phi(T,R) \Phi^\dagger(T^\prime, R^\prime)\rangle.
	\end{equation}
For $\omega$, $\omega^\prime > 0$ the Fourier transform with respect to $T$ and $T'$ at the spatial origin, $R=R'=0$, yields
	\begin{equation}\label{eq:Fouriertransf}
		\fl\int^\infty_{-\infty}\int^\infty_{-\infty} \frac{\mathrm{d} T \mathrm{d} T^\prime}{2 \pi} \ee^{-\ii \omega T} \mathcal{W}_0(T,0; T^\prime, 0) \ee^{\ii  \omega^\prime T^\prime} =  2 \pi F_{\omega}^{(e)*}(0,0) F_{\omega^\prime}^{(e)}(0,0) \langle b^{(e)\dagger}_{\omega} b^{(e)}_{\omega^\prime} \rangle.
\end{equation}	 
	
A manifestation of the Gibbons-Hawking effect is that the double Fourier transform~(\ref{eq:Fouriertransf}) of the two-point function in the Bunch-Davies vacuum state is that for the thermal state at inverse temperature $2\pi$:
\begin{equation}\label{eq:thermal}
		\langle b^{(e)\dagger}_{\omega} b^{(e)}_{\omega^\prime} \rangle = \frac{1}{\ee^{2 \pi \omega} - 1} \delta(\omega-\omega^\prime)\eqend{.} 
\end{equation}
By substituting this equation into (\ref{eq:Fouriertransf}) with $F_\omega^{(e)}(0,0)$ found from (\ref{eq:staticmodes1}) and (\ref{eq:coeff-A-even})  we find that the Gibbons-Hawking effect should lead to
\begin{equation}\label{eq:gibbons-hawking-result}
\fl		\eqalign{\int^\infty_{-\infty} \frac{\mathrm{d} T \mathrm{d} T^\prime}{2 \pi} \ee^{-\ii \omega T} \mathcal{W}_0(T,0; T^\prime, 0) \ee^{\ii  \omega^\prime T^\prime} &=  \frac{\ee^{- \pi \omega}}{8 \pi^2} \left\vert \Gamma \left( \frac{\ii \omega - l}{2} \right) \Gamma \left( \frac{1 + \ii \omega + l}{2} \right) \right\vert^2 \\ & \qquad \times \delta(\omega - \omega^\prime)\eqend{.}}
	\end{equation}
We now show this result for the two-point function for the Bunch-Davies vacuum state given by (\ref{eq:BD-2pt-function}).  Since the time coordinate for the global and static coordinates agree at $R=0$, we have
	\begin{equation}
		\mathcal{W}_0(T,0;T^\prime,0) = - \frac{1}{4 \sin l\pi} \mathsf{P}_l \left(- \cosh (T - T^\prime - \ii \epsilon) \right).
	\end{equation}
With the definition $\Delta T = T - T^\prime$, the Fourier transform of this two-point function is
	\begin{equation}\label{eq:Fouriertransf2}
		\eqalign{\int^\infty_{-\infty}\int^\infty_{-\infty} \frac{\mathrm{d} T \mathrm{d} T^\prime}{2 \pi} \ee^{- \ii \omega T} \mathcal{W}_0(T,0;T^\prime,0) \ee^{\ii  \omega^\prime T^\prime} \\ \quad = - \frac{1}{4 \sin l \pi} \delta(\omega - \omega^\prime) \int^\infty_{- \infty} \mathrm{d} \Delta T \ \ee^{- \ii \omega \Delta T} \mathsf{P}_l(- \cosh(\Delta T - \ii \epsilon)).}
	\end{equation}
For large $|z|$ the Legendre function $\mathsf{P}_l(z)$
is the sum of two terms, both decaying like $|z|^{-1/2}$ if $l = -1/2+\ii\lambda$, $\lambda\in\mathbb{R}$, and one decaying like $|z|^l$ and the other like $|z|^{-l-1}$ if $-1/2 < l < 0$  (see, e.g., \cite[Eq.~8.772.1]{gradshteyn2014table}).  Hence the function $\mathsf{P}_l(-\cosh(\Delta T-\ii\epsilon))$ decays exponentially fast for large $\Delta T$. Since the singularities of $\mathsf{P}_l(-\cosh(\Delta T-\ii \epsilon))$ are at $\cosh(\Delta T-\ii \epsilon) = 1$, i.e.\ at $\Delta T = \ii(2\pi n + \epsilon)$, $n\in\mathbb{Z}$, the integration path can be changed from the one from $-\infty$ to $\infty$ to the one from $-\infty-\ii\pi$ to $\infty - \ii\pi$.  Thus, letting $\Delta T = \tau - \ii\pi$, we find
\begin{eqnarray}
\fl \int_{-\infty}^\infty \D\Delta T\, \ee^{-\ii\omega \Delta T}\mathsf{P}_l(- \cosh(\Delta T-\ii\epsilon)) 
 & = & - \frac{\sin l \pi}{2 \pi^2}\ee^{-\pi\omega} \left\vert \Gamma \left( \frac{\ii \omega - l}{2} \right) \Gamma \left( \frac{1 + \ii \omega + l}{2} \right) \right\vert^2\eqend{,}\nonumber \\
\end{eqnarray}
where we have used the integral in~\cite[Eq.~7.165]{gradshteyn2014table}.
Substituting this expression into (\ref{eq:Fouriertransf2}), we recover (\ref{eq:gibbons-hawking-result}), which is a manifestation of the Gibbons-Hawking effect.

\section{Bound on an integral of the Ferrers function}\label{appendix:bound-Ferrers}

In this appendix we show that
if $l\in -1/2+\ii\mathbb{R}$ or $-1/2 < l < 0$, then the integral
\begin{equation}\label{eq:the-integral-of-P}
    I(\omega,s) = \int_{-\infty}^\infty \D t\, \mathsf{P}_l^{-s}(\ii \sinh t)\ee^{-\ii\omega t}\eqend{,}
\end{equation}
is bounded as
\begin{equation}\label{eq:bound-to-be-found}
    |I(\omega,s)| \leq C(\gamma)\ee^{-\pi\omega/2}\cosh(\pi u/2)\frac{|\Gamma(s+\frac{1}{2})|}
{\left|\Gamma(l+s+1)\Gamma(s-l)\right|}\eqend{,}
\end{equation}
with $s = \gamma+\ii u$, $\gamma \geq 0$, $u\in\mathbb{R}$, where $C(\gamma)$ is a positive constant independent of $\omega$ and $u$. 

First we change the argument of the Ferrers function in (\ref{eq:the-integral-of-P}) to $\cosh t$ as follows.  Since $\mathsf{P}_l^{-s}(z)$ is a linear combination of terms behaving like $z^l$ or $z^{-l-1}$ for large $|z|$, the integrand of (\ref{eq:the-integral-of-P}) tends to zero exponentially as $|t|\to \infty$ for the values of $l$ we are interested in.  This allows us to change the variable as $t\mapsto t-\ii\alpha$ as long as the integrand is non-singular between the real line and the line of constant imaginary part $-\ii\alpha$.  Since the singularities of $\mathsf{P}_l^{-s}(z)$ are at $z=\pm 1$ and since
\begin{equation}
\ii\sinh (t-\ii\alpha) = \ii\cos\alpha\sinh t + \sin\alpha \cosh t\eqend{,}
\end{equation}
we can choose $\alpha = \pi/2 - \epsilon$, where $\epsilon>0$ is infinitesimal with 
\begin{equation}
    \ii\sinh (t-\ii\pi/2+\ii\epsilon) = \cosh t + \ii\epsilon t\eqend{.}
\end{equation}
With this change of variable, we find
\begin{equation}
\mathsf{P}_l^{-s}(\ii\sinh t)  \mapsto  \mathsf{P}_l^{-s}(\cosh t+\ii\epsilon t) = 
\Big{\{}\begin{array}{ll} e^{\ii s\pi/2}P_l^{-s}(\cosh t) & \textrm{if}\ \ t < 0\eqend{,}\\
e^{-\ii s\pi/2}P_l^{-s}(\cosh t) & \textrm{if}\ \ t > 0\eqend{,} 
\end{array}
\end{equation}
where
\begin{eqnarray}
P_l^{-s}(z) & = & \left( \frac{z-1}{z+1}\right)^{s/2}
\frac{1}{\Gamma(1+s)}F(-l,l+1;1+s;(1-z)/2)\eqend{.}
\end{eqnarray}
Hence we find
\begin{equation}
I(\omega,s)  =  \ee^{-\pi \omega/2}
\int_0^\infty P_l^{-s}(\cosh t)\left( \ee^{-\ii s\pi/2} \ee^{-\ii\omega t} - \ee^{\ii s\pi/2}\ee^{\ii\omega t}\right)\Dt\eqend{,}
\end{equation}
and
\begin{equation}\label{eq:first-bound}
|I(\omega,s)| \leq 2\ee^{-\pi\omega/2}\cosh(\pi u/2)
\int_0^\infty |P_l^{-s}(\cosh t)|\Dt\eqend{,}
\end{equation}
where $s=\gamma+\ii u$, $\gamma \geq 0$.

Next we bound the integral of $|P_l^{-s}(\cosh t)|$ in (\ref{eq:first-bound}).  The formula~\cite[Eq.~8.713.3]{gradshteyn2014table} reads
\begin{equation}
P_\nu^{-\mu}(z) = \sqrt{\frac{2}{\pi}}\frac{\Gamma(\mu+\frac{1}{2})(z^2-1)^{\mu/2}}
{\Gamma(\nu+\mu+1)\Gamma(\mu-\nu)}
\int_0^\infty \frac{\cosh(\nu+\frac{1}{2})t\,\Dt}{(z+\cosh t)^{\mu+1/2}}\eqend{.}
\end{equation}
Hence
\begin{equation}\label{eq:bound-for-Pls}
\eqalign{\int_{0}^\infty |P_l^{-s}(\cosh t)|\Dt  & \leq  
 \sqrt{\frac{2}{\pi}} \left|\frac{\Gamma(s+\frac{1}{2})}
{\Gamma(l+s+1)\Gamma(s-l)}\right|  \\
& \quad \times \int_0^\infty \Dt \int_0^\infty \D v
\left|\frac{(\sinh t)^s\cosh(l+\frac{1}{2})v}{(\cosh t + \cosh v)^{s+1/2}}\right|\eqend{.}
}
\end{equation}
Recalling $-1/2 \leq \mathrm{Re}\,l < 0$ and choosing a constant $a\in (0,-\mathrm{Re}\,l)$, we find for $s=\gamma+\ii u$, $\gamma \geq 0$,
%
 \begin{equation}
     \fl \int_0^\infty \Dt \int_0^\infty \D v
\left| \frac{(\sinh t)^s\cosh(l+\frac{1}{2})v}{(\cosh t + \cosh v)^{s+1/2}}\right|
\leq \int_0^\infty \frac{|\sinh t|^\gamma}{|\cosh t|^{\gamma+a}}\Dt\int_0^\infty \frac{|\cosh(l+\frac{1}{2})v|}
{|\cosh v|^{\frac{1}{2}-a}}\D v\eqend{.}
 \end{equation}
 The integrals on the right-hand side of this inequality are convergent and independent of $u$.  Hence, using (\ref{eq:bound-for-Pls}) and (\ref{eq:first-bound}), we obtain the bound~(\ref{eq:bound-to-be-found}).

\section*{References}
\bibliographystyle{unsrt}

\bibliography{dS2ScalarsBib}

\end{document}